\documentclass[twocolumn]{aastex63}

\usepackage{amsmath}
\usepackage{amssymb}
\usepackage{bm} % italic bold math symbols

\let\tablenum\relax
\usepackage{siunitx}

\newcommand{\ketju}{{\normalfont \textsc{Ketju}}}
\newcommand{\epsPN}{\epsilon_\text{PN}}
\newcommand{\PN}[1]{\text{PN#1}}
\newcommand{\archain}{AR-CHAIN}
\newcommand{\gadget}{GADGET-3}

\newcommand{\dd}[1]{\mathop{}\!\mathrm{d}#1}
\newcommand{\dv}[2]{\frac{\dd#1}{\dd{#2}}} %derivative
\newcommand{\dvi}[2]{{\dd#1}/{\dd{#2}}} %inline version
\newcommand{\order}[1]{\mathcal{O}(#1)}
\newcommand{\abs}[1]{\left|#1 \right|}
\newcommand{\vc}[1]{\bm{#1}} %vectors

\newcommand{\eval}[1]{\left.#1\right\rvert}

\turnoffediting

\graphicspath{{figures/}}

\begin{document}
\title{
Gravitational Waves from the
Inspiral of Supermassive Black Holes in Galactic-scale Simulations
}
\shorttitle{GWs from SMBHs in Galactic Simulations}

\author[0000-0001-5721-9335]{Matias Mannerkoski}
\affiliation{
Department of Physics,
Gustaf H\"allstr\"omin katu 2, FI-00014, University of Helsinki, Finland
}

\author[0000-0001-8741-8263]{Peter H. Johansson}
\affiliation{
Department of Physics,
Gustaf H\"allstr\"omin katu 2, FI-00014, University of Helsinki, Finland
}

\author[0000-0003-1758-1908]{Pauli Pihajoki}
\affiliation{
Department of Physics,
Gustaf H\"allstr\"omin katu 2, FI-00014, University of Helsinki, Finland
}

\author[0000-0001-8789-2571]{Antti Rantala}
\affiliation{
Department of Physics,
Gustaf H\"allstr\"omin katu 2, FI-00014, University of Helsinki, Finland
}
\affiliation{Max-Planck-Institut f\"ur Astrophysik,
Karl-Schwarzchild-Str. 1,
D-85748, Garching, Germany
}

\author[0000-0002-7314-2558]{Thorsten Naab}
\affiliation{Max-Planck-Institut f\"ur Astrophysik,
Karl-Schwarzchild-Str. 1,
D-85748, Garching, Germany
}

\correspondingauthor{Matias Mannerkoski}
\email{matias.mannerkoski@helsinki.fi}

%\submitjournal{\apj}

\begin{abstract}
We study the orbital evolution and gravitational wave (GW) emission of
supermassive black hole (SMBH) binaries formed in gas-free mergers of massive early-type galaxies 
using the hybrid tree--regularized N-body code \ketju{}.
The evolution of the SMBHs and the surrounding galaxies is followed self-consistently
from the large-scale merger down to the final few orbits before the black holes coalesce.
Post-Newtonian corrections are included up to \PN{3.5}-level for the binary dynamics,
and the GW calculations include the corresponding corrections up to \PN{1.0}-level.
We analyze the significance of the stellar environment on the evolution of the binary
and the emitted GW signal during the final GW emission dominated phase of the binary hardening and inspiral.
Our simulations are compared to semi-analytic models that have often been used for making 
predictions for the stochastic GW background emitted by SMBHs.
We find that the commonly used semi-analytic parameter values produce large differences
in merger timescales and eccentricity evolution,
but result in only $\sim 10\%$ differences in the GW spectrum emitted by a single binary at frequencies 
$f\gtrsim 10^{-1} \, \rm yr^{-1}$, which are accessible by current pulsar timing arrays.
These differences are in part caused by the strong effects of the SMBH binaries on the 
surrounding stellar population, which are not included in the semi-analytic models.

\end{abstract}

\section{Introduction}

Supermassive black holes (SMBHs) with masses in the range  $M=10^{6} \text{--} 10^{10} M_{\sun}$ 
are believed to reside in the centers of most, if not all, massive galaxies
\citep[for a review, see][]{kormendy2013}. 
There is also increasing observational evidence for systems with
multiple SMBHs, such as an 
SMBH binary system at $z=0.055$ with a projected separation of just 
$\sim \SI{7}{pc}$ \citep{2006ApJ...646...49R,2017ApJ...843...14B} and a triple 
SMBH at $z=0.39$ with the closest pair separated by 
$\sim \SI{140}{pc}$ \citep{2014Natur.511...57D}.
The observed quasi-periodic outbursts of the quasar OJ287 have also plausibly 
been modeled as a binary system with a separation of only $\sim \SI{0.05}{pc}$ 
\citep{2008Natur.452..851V,2018ApJ...866...11D}.

In the $\Lambda$CDM hierarchical picture of structure formation, galaxies grow
through mergers and gas accretion, resulting in situations
with multiple black holes in the same galaxy 
\edit1{
\citep[e.g.][]{begelman1980, volonteri2003, 2018MNRAS.475.4967T}.
}
Galaxy mergers are particularly relevant for the most massive,
slowly-rotating early-type galaxy population hosting the largest SMBHs in the
Universe.
These galaxies are believed to have assembled through a two-stage
process in which the early assembly is dominated by rapid in situ star formation
fueled by cold gas flows and hierarchical merging of multiple star-bursting
progenitors, whereas the later growth below redshifts of $z\lesssim 2 \text{--} 3$ is
dominated by a more quiescent phase of accretion of stars brought in by minor
mergers
\citep[e.g.][]{2009Naab,oser2010,2012Johansson,2015Wellons,2017MNRAS.465..722F,2017ARA&A..55...59N,2018MNRAS.477.1822M}. 

Pioneering work by \citet{begelman1980} outlined the merging of SMBHs as a three-stage process. 
On larger scales the SMBHs are brought together through dynamical friction from stars and gas until 
a gravitationally bound hard binary with a semi-major axis of $a \sim \SI{10}{pc}$ is formed
in the center of the merging galaxy pair.
As the SMBH binary continues hardening stars become the primary scatterers,
experiencing complex three-body interactions 
that carry away energy and angular momentum from the SMBH binary system 
\citep[e.g.][]{1980AJ.....85.1281H}.
The largest uncertainty in this process is the rate at which the `loss cone' is filled,
i.e. the region of parameter space where the stars have
sufficiently low angular momenta to interact with the SMBH binary.
If the SMBH `loss cone' is depleted, the binary hardening 
is halted and we are faced by the so-called final-parsec problem \citep{Milosavljevic2001,merritt2013}.
However, if the SMBH binaries are able to reach
smaller separations $(\lesssim \SI{0.1}{pc})$ either aided by a repopulation of the loss cone 
\citep[e.g.][]{berczik2006,2013ApJ...773..100K,Vasiliev2015,2017MNRAS.464.2301G}
and/or gas drag \citep[e.g.][]{Mayer2007,Chapon2013,2016ApJ...828...73K},    
the loss of orbital energy eventually becomes dominated by the emission of 
gravitational waves at very small centiparsec binary separations \citep{1963PhRv..131..435P}. 

The recent direct detection of gravitational waves (GWs) by the LIGO scientific 
collaboration confirmed observationally the GW driven  merger scenario for stellar
mass BH binary systems \citep{Abbott2016}.
However, the expected frequencies  of gravitational waves from binary SMBHs are 
several orders of magnitude lower and hence undetectable by LIGO, or 
any other ground-based interferometer.
Instead the direct detection of GWs from SMBH binaries has to wait for 
planned space-based missions, such as LISA (Laser Interferometer Space Antenna),
which will be sensitive at  sufficiently low frequencies $(f=\num{e-4} \text{--} \SI{e-1}{Hz})$
and thus able to detect the final stages of the inspiral 
and coalescence of SMBHs with masses of $M_{\bullet}\lesssim 10^8 M_\sun$ \citep{2017arXiv170200786A}. 
  
In the meantime before LISA becomes operational, pulsar timing arrays (PTAs) constitute
the most promising method for detecting gravitational waves emitted by SMBH binaries.
PTAs attempt to detect GWs at nanohertz frequencies by measuring correlated
offsets in the arrival times of the highly regular pulses emitted by pulsars in 
the Milky Way (see e.g. \citealt{2018PASA...35...13T} for an overview).
It is expected that the large population of 
GW sources forms a stochastic gravitational wave background (GWB), which has a nearly powerlaw spectrum \citep{2001astro.ph..8028P} 
with a characteristic amplitude of $h_c \sim 10^{-15}$ at the reference frequency of $f=\SI{1}{yr^{-1}} \approx \SI{3e-8}{Hz}$.
The exact properties of the GW spectrum are affected by the eccentricity distribution of the binaries and different environmental effects, such as
stellar `loss-cone' scattering and the viscous drag from circumbinary gas discs that will influence the binary population 
\citep[e.g.][]{2013CQGra..30v4014S,2017MNRAS.471.4508K,2019A&ARv..27....5B}.
The PTA observations hold great promise for 
constraining the properties and formation mechanisms of the SMBH binary population  
\citep[e.g.][]{2018ApJ...859...47A, 2018NatCo...9..573M, 2019MNRAS.488..401C}.

Previous work on the GW emission from merging SMBH binaries can be
classified into two categories based on whether the final sub-parsec dynamics of
the binary is directly resolved or instead modeled using semi-analytic formulae.
Studies of the GWB amplitude are typically in the latter category, with
the evolution due to gravitational wave emission implemented often using the classic
leading order formulae of \citet{1964PhRv..136.1224P}. 
Effects of the environment may be included using simplified models which cannot 
accurately account for e.g. highly anisotropic stellar populations.
The main differences between the different 
semi-analytic studies lie in the treatment of the population of binaries, using either analytic formulae
constrained by observations of galaxies \citep[e.g.][]{2014ApJ...789..156M, 2015PhRvD..92f3010H, 2018ApJ...863L..36I} 
and observed candidate binary systems \citep{2018ApJ...856...42S}, 
or the results from cosmological simulations 
\citep{2008MNRAS.390..192S, 2016MNRAS.463..870S, 2017MNRAS.471.4508K,2018MNRAS.473.3410R}.

Studies where the dynamics of the SMBH binary is resolved down to the scales
relevant for GW emission either model the environment using semi-analytic
methods \citep[e.g.][]{2016MNRAS.461.4419B, 2018MNRAS.477.2599B},
or they utilize N-body codes to directly simulate the
interaction between the binary and the surrounding field of stars, using either
Newtonian gravity \citep{2011ApJ...732L..26P} or including also post-Newtonian
(PN) corrections to the motion of the SMBHs to account for relativistic effects
\citep{2009ApJ...695..455B, 2018A&A...615A..71K}.
Typically, due to the steep $\order{N^2}$ scaling of computational time with particle number,
these simulations are limited to a relatively low number of
particles, which typically restricts them to studying only the core regions of
the galaxies surrounding the SMBHs. Some of these resolution issues can be mitigated by using a combination 
of codes for different phases of the galaxy merger and SMBH inspiral 
\citep{2016ApJ...828...73K,2018ApJ...868...97K}.

In this paper we utilize the recently developed hybrid tree--N-body code \ketju{}
\citep{2017ApJ...840...53R,2018ApJ...864..113R} to directly compute the trajectories of the SMBHs and the resulting 
GW signal in gas-free mergers of massive early-type galaxies. 
The \ketju{} code uses an algorithmic chain regularization (\archain{}); 
\citep{Mikkola2006,Mikkola2008} method to efficiently and accurately compute the dynamics close to SMBHs,
and combines it with
the fast and widely used tree code \gadget{} 
(\citealt{Springel2005}; see also \citealt{2015MNRAS.452.2337K} for a similar code combination).
The hybrid nature of \ketju{} enables us to follow the SMBHs from before the galaxies merge until the final
few orbits before the coalescence of the SMBH binary that forms after the
merger. In particular, the distribution of stars is followed self-consistently throughout the simulation,
correctly accounting for the changing properties of the surrounding stellar population during
both the dynamical friction dominated phase and the stellar scattering driven phase.

The main goal of the present work is to compare the gravitational wave emission from 
dynamically resolved binary SMBHs in \ketju{} simulations to semi-analytic models that have been
used to compute the GW emission for SMBH binaries in e.g. cosmological
simulations, where the spatial resolution is insufficient to model the SMBHs directly
\citep{kelley2017,2017MNRAS.471.4508K}.
Our main focus for the GW calculations is in the PTA frequency band
and the comparison with semi-analytic models allows us to
validate the semi-analytic models and directly address  
how important the additional accuracy gained by using \ketju{} is
in the context of making predictions for PTA observations.

The remainder of this paper is organized as follows: 
in \autoref{sec:simulations} we give a brief overview of the \ketju{} code and
describe our merger simulations.
In \autoref{sec:orbit_analysis} we present the methods used for analyzing the
evolution of the SMBH binary, including the use of quasi-Keplerian orbital elements
that account for post-Newtonian corrections as well as different simple binary evolution models
used for comparison.
Then, in \autoref{sec:gw_computation}, we present the combination of two 
methods used to compute the emitted gravitational waves in different phases of
the binary evolution.
These methods are applied to the simulations in \autoref{sec:results}, and the
implications of the obtained results are discussed in \autoref{sec:discussion}.
Finally, we present our conclusions in \autoref{sec:conclusions}.

\section{Numerical Simulations}
\label{sec:simulations}

\begin{figure*}
\plotone{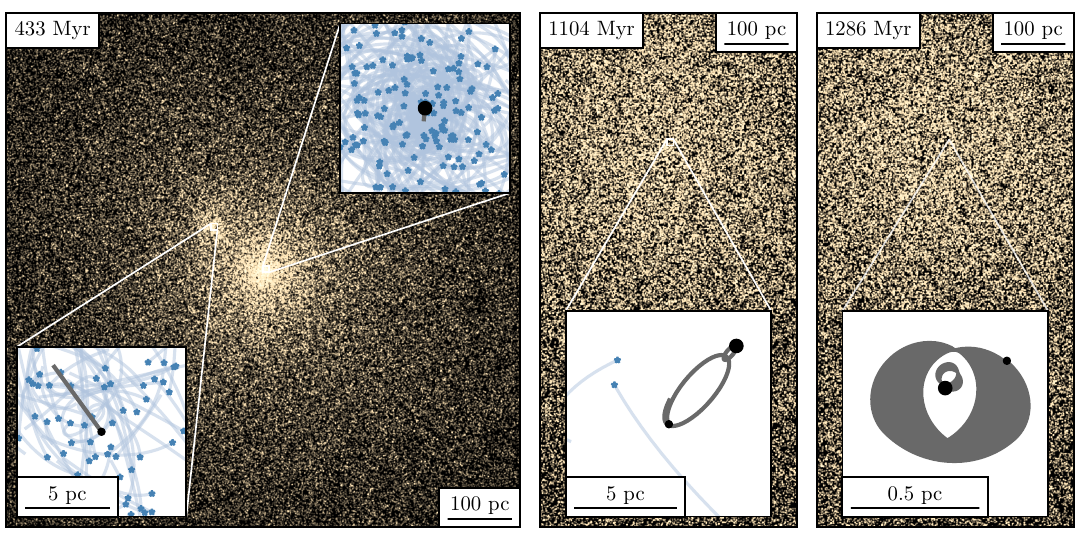}
\caption{
A sequence of illustrative example snapshots of simulation run A (5:1 mass ratio merger).
The main images show the projected stellar density, while the insets show the 
stellar particles (blue), SMBHs (black), and sections of their trajectories 
in the regularized region (grey).
The snapshots illustrate different characteristic phases of the SMBH binary evolution:
the SMBHs first sink to the center of the merging galaxy due to dynamical friction (left panel)
and form an eccentric bound binary (middle panel) that shrinks due to stellar scattering,
finally entering the strongly relativistic regime (right panel) where the binary shrinks
due to GW emission, and other relativistic effects such as precession of the orbit are apparent as well.
The corresponding timescales and spatial scales are indicated on the figure. 
}
\label{fig:prplot}
\end{figure*}

\subsection{The KETJU Code}

The simulations analyzed in this paper have been run using the recently developed 
\ketju{} code (see \citealt{2017ApJ...840...53R,2018ApJ...864..113R}  for full code details), which is an extension of the  
tree-SPH simulation code \gadget{} \citep{Springel2005}. The central idea of the code is the 
inclusion of a regularized region around every SMBH particle, in which the non-softened gravitational dynamics
is computed  using the regularized \archain{} \citep{Mikkola2008} integrator while the dynamics of the 
remaining particles is computed with the \gadget{} leapfrog using the tree force calculation method.

In practice the code operates by dividing simulation particles into three categories. The SMBH and all stellar 
particles, which lie within a user defined chain radius $(r_{\rm chain})$ are marked 
as chain subsystem particles. Particles that lie just outside the chain radius, 
but induce a strong tidal perturbation on the chain system 
are marked as perturber particles. Finally, all the remaining particles that are 
far from any SMBHs are treated as ordinary \gadget{} particles 
with respect to the force calculation. 
The \ketju{} code also allows for both multiple simultaneous chain subsystems 
and several SMBHs in a given single subsystem.

During each global \gadget{} timestep the particles in the chain subsystems are 
propagated using the \archain{} algorithm \citep{Mikkola2008,2017ApJ...840...53R}.
This algorithm has three main aspects: algorithmic regularization, the use of 
relative distances to reduce round-off errors by organizing the particles in a chain 
(e.g. \citealt{Mikkola1993}) and finally the use of a Gragg--Bulirsch--Stoer (GBS);
\citep{Gragg1965,Bulirsch1966} extrapolation method that yields 
high numerical accuracy in orbit integrations at a preset user-given error 
tolerance level $(\eta_\mathrm{GBS})$. In essence, algorithmic regularization 
works by transforming the equations of motion by introducing a fictitious time variable such that 
integration by the common leapfrog method yields 
exact orbits for a Newtonian two-body problem including two-body collisions
\citep[e.g.][]{1999MNRAS.310..745M, 1999AJ....118.2532P}.

\subsection{Post-Newtonian Corrections}

The \archain{} algorithm within the \ketju{} code features an extension of  
phase space with the help of an auxiliary velocity variable \citep{Hellstrom2010, Pihajoki2015}.
This allows for the efficient implementation of the 
velocity-dependent post-Newtonian corrections in the motion of the SMBH 
particles \citep[e.g.][]{Will2006} using an explicit integrator. 

Schematically the PN-corrected acceleration can be written as
\begin{equation}
\vc{a} = \vc{a}_\text{N} + 
\vc{a}_\text{\PN{1}} + 
\vc{a}_\text{\PN{2}} + 
\vc{a}_\text{\PN{3}} + 
\vc{a}_\text{\PN{2.5}} + 
\vc{a}_\text{\PN{3.5}}, 
\end{equation}
where the Newtonian acceleration $\vc{a}_\text{N}$ is computed including the
surrounding stellar particles, while the PN-terms only include contributions
from other SMBHs.
The PN-correction terms are labeled so that they are proportional to the 
corresponding power
of the formal PN expansion parameter $\epsPN$, i.e.
\begin{equation}
\abs{\vc{a}_{i \text{PN}}} \propto \epsPN^{i} 
\sim \left(\frac{v}{c}\right)^{2i} 
\sim \left( \frac{R_s}{R}\right)^i,
\end{equation}
where $v$ and $R$ are the relative velocity and separation of a
pair of SMBHs, $R_s = 2 GM/c^2$ is the Schwarzschild radius corresponding to the
total binary mass $M$, $c$ is the speed of light and $G$ the gravitational constant.
The PN terms of integer order are conservative and are associated with
conserved energy and angular momentum, while the half integer order terms are
dissipative radiation reaction terms caused by the emission of gravitational radiation.

In the \ketju{} simulations studied in this paper we use PN-correction terms up to order \PN{3.5} derived for a binary
system \edit1{of arbitrary eccentricity} in the modified harmonic gauge
\edit1{as given in} \cite{2004PhRvD..69j4021M}.
Terms depending on the spin of the black holes as well as the lowest order cross
term corrections for a system consisting of more than two particles have also
been implemented in the \ketju{} code, but are not used in this present work.
The spin terms are ignored due to the fact that the spin of the SMBHs is largely
determined by interactions with gas, which is not included in this study.
Similarly, the cross terms as well as other PN-corrections for stellar particles
are ignored due to the unphysically strong
corrections resulting from stellar particles with very large masses $(m_\star \gg M_\sun)$.
\edit1{Thus post-Newtonian corrections are only included in the interactions
between SMBHs and only when they are within the same chain region.}

We note that the missing higher order PN-corrections may potentially lead to significant
effects over long periods of time.
\edit1{
Corrections up to the \PN{4}-level have been recently derived
\citep[e.g.][]{2018PhRvD..97d4037B}, but these are highly impractical to implement due
to the appearance of the so called `tail'-terms that depend on the entire
history of the binary in addition to its instantaneous state.
}
The lack of these higher order terms \edit1{also} leads to the simulated motion of the SMBHs
becoming unreliable when $\epsPN \sim 0.1$ \citep[e.g.][]{2012CQGra..29x5002C}.
This is most apparent when considering the conservation of energy in the system,
which we will discuss at the end of \autoref{sec:results}.
\edit1{
Due to the increasing unreliability of the PN-corrections, 
we stop simulating the binaries at a separation of $R = 6 R_s$,
where the equations of motion are still well behaved.
However, as will be discussed in \autoref{sec:results},
the actual limit of reliability for our purposes is closer to $R \sim 10 R_s$.
}

\subsection{Progenitor galaxies and merger orbits}
\label{sec:progenitors}

\begin{deluxetable*}{ccccccccccccc}
\tablecaption{
Summary of the studied merger runs.
\label{table:runs}
}
\tablehead{
\colhead{Run} & \colhead{$m_{\bullet 1}$} & \colhead{$m_{\bullet 2}$} 
& \colhead{$M_\bullet$} 
& \colhead{$M_\text{DM}$} & \colhead{$M_\star$}
& \colhead{$N_\text{DM}$} & \colhead{$N_\star$} 
&\colhead{$q$} & \colhead{$a_0$} & \colhead{$e_0$} 
& \colhead{$t_0$} & \colhead{$t_\text{merger}$}
\\
& \colhead{$(10^9 M_\sun)$} & \colhead{$(10^9 M_\sun)$} 
& \colhead{$(10^{10} M_\sun)$}
& \colhead{$(10^{13} M_\sun)$} & \colhead{$(10^{10} M_\sun)$}
& \colhead{$(\times 10^6)$} & \colhead{$(\times 10^6)$}
& & \colhead{(pc)} &
& \colhead{(Gyr)} & \colhead{(Gyr)}
}
\startdata
A & 8.5  & 1.7 & 1.02 & 9.0   & 49.8   & 12.0  & 4.98    &  5:1 & 4.93  & 0.958 & 0.979 & 1.30 \\
B & 10.2 & 1.7 & 1.19 & 10.5  & 58.1   & 14.0  & 5.81    &  6:1 & 5.45  & 0.971 & 1.09  & 1.22 \\
C & 11.9 & 1.7 & 1.36 & 12.0  & 66.4   & 16.0  & 6.64    &  7:1 & 4.68  & 0.961 & 1.37  & 1.50 \\
D & 13.6 & 1.7 & 1.53 & 13.5  & 74.7   & 18.0  & 7.47    &  8:1 & 5.32  & 0.954 & 2.10  & 2.38 \\
X & 0.4  & 0.4 & 0.08 & 1.942 & 16.82  & 19.42 & 1.682   &  1:1 & 0.520 & 0.925 & 0.839 & 1.05 \\
\enddata

\tablecomments{
Listed are the masses of the individual SMBHs ($m_{\bullet 1}$, $m_{\bullet 2}$),
the total binary mass ($M_\bullet$), the total dark matter and stellar masses and particle numbers of the merger remnants
($M_{\text{DM},\star}$, $N_{\text{DM},\star}$), the merger mass ratio $q$, 
the SMBH binary orbital parameters $a_0$ and $e_0$ at the time $t_0$,
as well as the time of the SMBH binary merger $t_\text{merger}$.
The times are measured from the start of the simulation,
and the time period considered in this work spans the interval $(t_0, t_\text{merger})$.
}
\end{deluxetable*}

Our merger progenitor galaxies consist of 
spherically symmetric, isotropic multi-component systems 
consisting of a stellar bulge, a dark matter halo and a central SMBH. The stellar component 
follows a  Dehnen profile  \citep{Dehnen1993} with $\gamma = 3/2$, whereas the dark matter halo 
is modeled by a  Hernquist ($\gamma = 1$) profile \citep{hernquist1990}. The multi-component 
initial conditions are generated using the distribution function method (e.g. \citealt{Merritt1985}) following 
the approach of \citet{Hilz2012}. In this method the distribution functions $f_{i}$ 
for the different mass components are obtained from the corresponding density profile $\rho_{i}$ and the 
total gravitational potential $\Phi_{T}$ is then derived using Eddington's formula \citep{binney2008}. 

In this paper we primarily study the orbits and merging of SMBHs in unequal-mass mergers. The primary galaxy is identical 
to the `$\gamma$-1.5-BH-6' initial condition of \cite{2018ApJ...864..113R,2019ApJ...872L..17R}, which was used to model
the major merger progenitor of NGC 1600 \citep{thomas2016}. This model corresponds to a massive gas-free early-type galaxy with 
a stellar mass of $M_\star=4.15\times 10^{11} M_{\sun}$ with an effective radius of $R_\mathrm{e}= \SI{7}{kpc}$ and a dark matter
halo mass of $M_\mathrm{DM}=7.5 \times 10^{13} M_{\sun}$. The dark matter fraction within the effective radius is set to 
$f_\mathrm{DM}(R_\mathrm{e})=0.25$ and the central SMBH has a mass of $M_\bullet=8.5\times 10^{9} M_{\sun}$. The secondary galaxy 
is a scaled down version of the primary galaxy, with the masses of all components divided by a factor of 5, i.e. 
$M_\star=8.3\times 10^{10} M_{\sun}$, $M_\mathrm{DM}=1.5 \times 10^{13} M_{\sun}$,  $M_\bullet=1.7\times 10^{9} M_{\sun}$ and 
a resulting effective radius of $R_\mathrm{e}=3.5 \, \rm kpc$  (this corresponds to IC-1 in \citealt{2019ApJ...872L..17R}). 
For this simulation set the masses of stellar particles are set to $m_\star = 10^5 M_\sun$ and $m_\mathrm{DM} = 7.5\times10^6 M_\sun$
for the dark matter particles resulting in $N_\star=4.15\times 10^{6}$ stellar and $N_\mathrm{DM}=1.0 \times 10^{7}$ dark matter particles 
for the primary galaxy, the number of particles being a factor of five lower for the secondary galaxy. 

We first simulate a 5:1 minor merger between the primary galaxy (Simulation A in Table \ref{table:runs}).
Next we continue the 5:1 merger run with subsequent merger generations until a total of 
4 minor mergers are completed (Simulations B--D in Table \ref{table:runs}). The fifth merger generation simulation could not
be used for the present study, as the black holes did not merge within a reasonable simulation time due to the formation of a 
very low density core in this system, with the missing stellar mass effectively giving rise to a 
final-parsec problem. 

Finally, in addition to the unequal-mass mergers we simulated an equal-mass 
 major merger between two identical galaxies with significantly lower masses of 
$M_\star=8.41\times10^{10} M_{\sun}$, $M_\mathrm{DM}=9.71\times10^{12} M_{\sun}$,
$M_\bullet=4\times10^8 M_{\sun}$ and an effective radius of  $R_\mathrm{e}= \SI{4.0 }{kpc}$ (this corresponds to the IC in \citealt{Eisenreich2017} excluding the hot gas halo). 
For this run we used $N_\star=8.41\times10^5$ stellar and $N_\mathrm{DM}=9.71\times10^{6}$ dark matter particles for 
each galaxy, resulting in particle masses of $m_\star =10^5 M_{\sun}$ and $m_\mathrm{DM} = 10^6 M_{\sun}$ for the stellar and dark matter components, respectively and a final merged SMBH mass of $M_\bullet=8\times10^8 M_{\sun}$ (simulation X in Table \ref{table:runs}).
The motivation for this run was to study the effect of the stellar environment on the 
SMBH merging process in a setting where the black hole masses are lower by a factor of $\sim 10$ compared to the high SMBH mass A--D 
simulation set. 

All merger orbits are nearly parabolic, with the pericenter distance set to $r_\mathrm{p} \sim 0.5 R_\mathrm{e}$ 
of the primary galaxy. After each minor merger (runs A--D), the merger remnant is reoriented so that the satellite 
galaxies fall in from random directions with respect to the principal axis of the primary galaxy. 
In \autoref{fig:prplot} we show a sequence of illustrative snapshots of run A, depicting the evolution of the SMBH binary 
from the dynamical friction phase (left panel), through the stellar scattering stage (middle panel) 
and into the final gravitational wave driven inspiral phase (right panel).

\subsection{Modeling the Final Phases of SMBH Inspirals}
\label{sec:sim_runs}
Following \citet{2018ApJ...864..113R,2019ApJ...872L..17R} we set the \gadget{} integrator error 
tolerance to $\eta = 0.002$ and the force accuracy to $\alpha = 0.005$, using the standard cell opening criterion 
\citep{Springel2005}. The chain radius is set to $r_\mathrm{chain} = \SI{10}{pc}$ and the perturber radius 
to twice this value, i.e. $r_{\mathrm{pert}} = 2  r_{\mathrm{chain}}$, for all the simulation runs.  
The gravitational softening lengths are set to $\epsilon_\star = 3.5 \, \rm pc$ and $\epsilon_\mathrm{DM} = 100 \, \rm pc$, respectively. The softening lengths are chosen to fulfill the criterion $r_\mathrm{chain} > 2.8 \times 
\epsilon$ \citep{2017ApJ...840...53R}.
\edit1{In interactions between particles with different softening lengths
\gadget{} uses the larger softening length.}
 
The mergers are initially evolved using \ketju{} with a chosen GBS tolerance 
of $\eta_\mathrm{GBS}= 10^{-6}$ until the binary black hole semi-major axis is around $a \sim 5000 R_s$. % (Schwarzschild radii). 
This corresponds to about $a_{0}\sim \SI{5}{pc}$ for runs A--D and $a_{0}\sim \SI{0.5}{pc}$ for the lower mass run X, 
with all SMBH binaries having  high eccentricities, in excess of $e_0 \gtrsim 0.9$. These separations are typically reached 
within $t_{0}\sim 1\text{--}2 \, \si{Gyr}$ after the start of the simulation (see Table \ref{table:runs}). 

At this point of time, we restart our \ketju{} runs with a stricter GBS tolerance value of 
$\eta_\mathrm{GBS}= 10^{-9}$. The increased accuracy is required for  calculating the resulting GW spectrum and resolving in detail the post-Newtonian orbital motions of the individual SMBHs all the way down to their final coalescence, which typically occurs $\Delta t_{\rm merger}\sim 100\text{--}300 \, \rm Myr$ later (see Table \ref{table:runs}).    
\edit1{
The simulation of the binary in these runs ends when the estimated merger
timescale of the binary is smaller than the timestep used for the global
simulation. At this point the separation of the binary is typically 50--70
Schwarzschild radii.    
}

We use the SMBH positions and  velocities computed in the full \ketju{} runs that include the influence of the stellar background for our GW calculations until the separation of the SMBHs is  $\lesssim 100 R_s$.
\edit1{
This distance corresponds to $\sim \SI{0.1}{pc}$ for runs A--D, and to $\sim
\SI{8e-3}{pc}$ for run X.
}
The final part of the merger is then run separately without the surrounding stellar particles at a much higher time resolution to facilitate direct waveform
computations as described in section \ref{sec:gw_computation}.
\edit1{
The transition separation of $\sim 100 R_s$ was chosen as it provides a good
balance between computational cost and the accuracy of these calculations.
}
While in principle stellar particles could also be included in this phase, we find that ignoring the environment in this very final stage does not cause significant
changes in the SMBH evolution, as only very few stellar particles are close enough to influence the SMBH binary at our mass resolution. 
\edit1{
This has also been checked by comparing the evolution of the isolated binaries
to the results from the full \ketju{} run down to separations of $\sim 70 R_s$,
where the lower time resolution \ketju{} data are still available.
}

\section{SMBH binary orbit analysis methods}
\label{sec:orbit_analysis}

\subsection{Post-Newtonian Orbital Elements}

\begin{figure*}
\plottwo{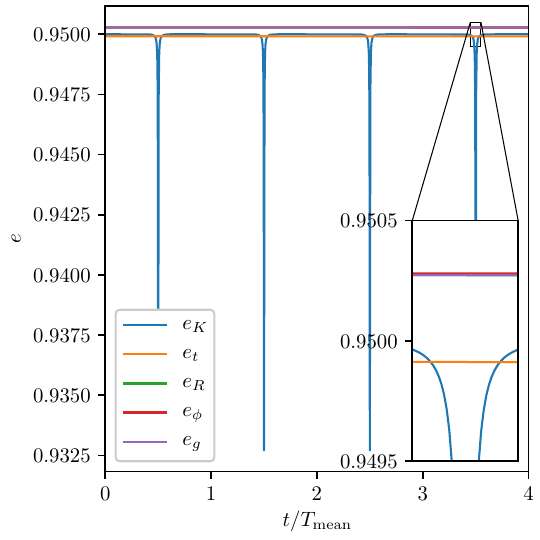}{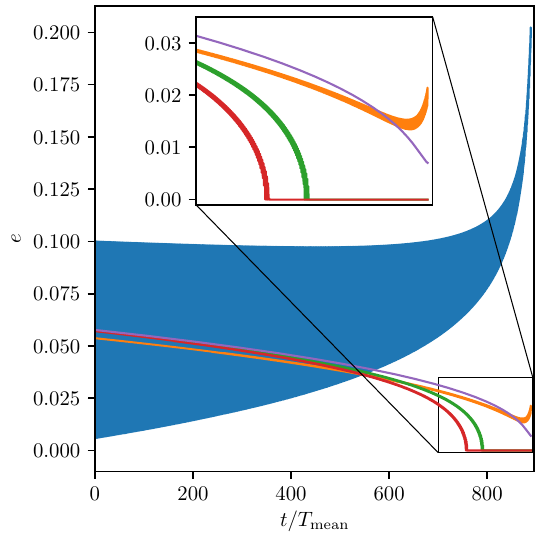}
\caption{
Comparison of the different eccentricity definitions for an isolated highly eccentric
binary with a semi-major axis of $a \approx 5000 R_s$ evolved for four orbital
periods using \ketju{} (left panel) and for a low
eccentricity binary evolved from $a \approx 25 R_s$ down to $a \approx 6 R_s$
over $\sim 1000$ orbital periods (right panel).
The time is given in units of the mean orbital period $T_\text{mean}$ over the
considered part of the orbital evolution.
The eccentricities shown are the Keplerian eccentricity $e_K$ computed from the
position and velocity using standard non-relativistic formulae; the
quasi-Keplerian eccentricities $e_t$, $e_R$, and $e_\phi$; and
the geometric eccentricity $e_g$ of equation \eqref{eq:geometric_eccentricity}.
All the line widths are equal, with the thicker appearance of some lines caused
by oscillations occurring during a single orbit.
In the left panel the geometric eccentricity overlaps with $e_R$ and $e_\phi$,
which differ slightly from $e_t$.
The poor behavior of the Keplerian eccentricity is evident in both cases.
}
\label{fig:eccentricity_comparison}
\end{figure*}

For an analysis of the evolution of the SMBH binary orbit, it is necessary to have
a parametrization of the orbit that changes slowly and can be recovered from the
instantaneous position and velocity of the binary.
Such a parametrization allows also computing the gravitational wave emission
of the system from only sparsely stored data points, significantly 
reducing the computational effort.

In Newtonian gravity the orbit of a binary system can be described by the
Keplerian orbital elements including the semi-major axis $a$ and the
eccentricity $e$, which are constant or evolve only slowly due to external
perturbations.
However, due to  the inclusion of PN-corrections to the motion of the SMBH binary, the
standard Keplerian elements are no longer even approximately constant over an orbit,
but rather oscillate quite strongly especially near the pericenter of an eccentric
orbit, 
as is illustrated in \autoref{fig:eccentricity_comparison}.

As a result of this, the Keplerian parameters computed from sparsely stored data
suffer from essentially random fluctuations due to the varying phase of the
orbit at the data points, which is problematic both for analyzing the evolution
of the orbit as well as for GW computations.
For small binary separations, the Keplerian orbital parameters are also no
longer related to the geometry of the orbit, as is apparent from the increase of
the Keplerian eccentricity for a nearly circular binary in the right panel of 
\autoref{fig:eccentricity_comparison}.

The problems associated with the Keplerian parameters can be mostly remedied by using
a quasi-Keplerian parametrization which accounts for the effects of the
PN-corrections.
Several such parametrizations accurate to different orders have been developed,
and
in this work we utilize the \PN{3} accurate quasi-Keplerian orbital parametrization
of \citet{2004PhRvD..70j4011M}.
This parametrization gives an approximate solution to the
conservative part of the \PN{3} accurate equations of motion of a non-spinning binary in the
form
\begin{equation} \label{eq:qkepler1}
R = a (1-e_R \cos{u}),
\end{equation}
\begin{equation} \label{eq:qkepler2}
\begin{aligned}
n (t-t_0) = u -& e_t \sin{u} 
             + \left( \frac{g_{4t}}{c^4} + \frac{g_{6t}}{c^6}\right)(v-u)\\
             +& \left( \frac{f_{4t}}{c^4} + \frac{f_{6t}}{c^6}\right)\sin{v}
             + \frac{i_{6t}}{c^6} \sin{2v} \\
             +& \frac{h_{6t}}{c^6} \sin{3v},
\end{aligned}
\end{equation}
\begin{equation} \label{eq:qkepler3}
\begin{aligned}
\frac{2\pi}{\Phi} (\phi-\phi_0) = v 
        +& \left( \frac{f_{4\phi}}{c^4} + \frac{f_{6\phi}}{c^6}\right)\sin{2v}\\
        +& \left( \frac{g_{4\phi}}{c^4} + \frac{g_{6\phi}}{c^6}\right)\sin{3v}
        + \frac{i_{6\phi}}{c^6} \sin{4v} \\
        +& \frac{h_{6\phi}}{c^6} \sin{5v},
\end{aligned}
\end{equation}
\begin{equation} \label{eq:qkepler4}
\tan{\frac{v}{2}} = \sqrt{\frac{1+e_\phi}{1-e_\phi}} \tan{\frac{u}{2}}. 
\end{equation}

The coordinates $R$ and $\phi$ are the relative separation and angle in the
orbital plane, with $\phi_0$ fixing the direction of pericenter at time $t=t_0$.
As in the Keplerian case, the orbit is an ellipse with semi-major axis $a$,
but now the frequency of radial oscillations $f_r = n/2\pi$ 
is not the same as the frequency of angular motion.
Rather, during a single radial period the direction angle $\phi$ traverses an
angle $\Phi>2\pi$ leading to the precession of the orbit.
The angles $u$ and $v$ are generalizations of the eccentric and true anomaly,
although their interpretation in terms of the geometry of the orbit is not
as straightforward as in the Keplerian case due to the appearance of the
additional series expansion factors $g,f,i$ and $h$
in the generalization of Kepler's
equation \eqref{eq:qkepler2} and in equation \eqref{eq:qkepler3}.
In addition the single eccentricity of the Keplerian parametrization is replaced
by three different eccentricities $e_R$, $e_t$ and $e_\phi$.
In the limit where the post-Newtonian corrections are negligible this
parametrization reduces to the standard Keplerian parametrization. 

The orbital parameters appearing in the above parametrization can be computed
from the binary separation vector $\vc{R}$ and relative velocity $\vc{V}$ in the
modified harmonic gauge used by \ketju{} via \PN{3} accurate approximations of the
conserved energy $E$ and angular momentum $J$ in the form
\begin{gather}
\label{eq:pn_energy_ang_momentum}
    E = \mu \left(E_0 + c^{-2} E_1 +c^{-4} E_2 + c^{-6} E_3\right)\\
    J = \mu \abs{\vc{R} \times \vc{V}} \left(J_0 + c^{-2} J_1 +c^{-4} J_2
   + c^{-6} J_3\right),
\end{gather}
where $\mu = m_1 m_2/M$ is the reduced mass and  $M = m_1+m_2$ the total mass of
the binary.
The expressions for the $E_i, J_i$ in terms of $\vc{R}$ and $\vc{V}$
are quite lengthy, and can be found in \citet{2004PhRvD..70j4011M}.
These approximations of the conserved energy and angular momentum are 
not constant along an orbit even when the radiation reaction terms are not
included.
Instead they oscillate slightly, and this oscillatory behavior carries over to
the orbital parameters, which are themselves given as \PN{3} accurate expressions
in terms of the approximated $E$ and $J$.
The inclusion of the radiation reaction terms also introduces additional periodic
oscillations \citep{2006PhRvD..73l4012K}, but
all of the oscillations are typically much smaller than the resulting secular
evolution, and thus of no significant consequence.

The appearance of three different eccentricities causes some ambiguity in
their interpretation, as they do not correspond to a single orbital ellipse as in the
Keplerian case.
The radial eccentricity $e_R$ appears to be most directly related to the
dimensions of the orbit, but unfortunately the \PN{3} formulae for it and $e_\phi$
fail for nearly circular relativistic orbits, resulting in $e_R^2<0$ and $e_\phi^2 < 0$.
The time eccentricity $e_t$ remains real, but it shows a small increase in this
regime which does not correspond to an actual increased eccentricity of the orbit.
A measure of eccentricity that describes the actual geometry of the orbit even
in the strongly relativistic regime can be defined as 
\citep[e.g.][]{2012CQGra..29x5002C}
\begin{equation} \label{eq:geometric_eccentricity}
e_g = \frac{R_\text{max} -R_\text{min}}{R_\text{max}
+R_\text{min}},
\end{equation}
where $R_\text{min}, R_\text{max}$ are the peri- and apocenter separations of
the binary.
This definition is impractical since it cannot be computed from the
instantaneous relative position and velocity of the binary, but it serves as a
useful comparison.

The right panel of \autoref{fig:eccentricity_comparison} shows the near-merger
behavior of the different eccentricities for an isolated binary with an
eccentricity that is comparable to the ones seen in the main \ketju{} runs at this 
separation.
The time eccentricity $e_t$ can be seen to be quite
close to the geometric eccentricity $e_g$ for the most part,
while both $e_R$ and $e_\phi$ go to zero too early, which is caused by them
becoming imaginary at that point.
In regimes where the PN expansion parameter $\epsPN$ is small the differences 
between the different eccentricities are negligible, so in the following 
we elect to use $e \equiv e_t$ as the single 
parameter describing the eccentricity of the orbit.
However, one should keep in mind that for the final stages of the binary
inspiral the eccentricity defined this way does not exactly describe the shape
of the orbit.
The other orbital elements likely suffer from similar deterioration in the very final
stages of the inspiral, but for the vast majority of the binary orbital evolution they
work reliably.

\subsection{Extracting the Effects of the Environment}
\label{sec:extracting_env}

The stellar environment affects the evolution of the binary orbit in addition to
the evolution caused by GW emission.
In the final stages of the binary evolution considered in this work these
effects are however fairly small, and to analyze them it is necessary to
separate them from the main GW-driven orbital evolution.
The time derivative of a given quantity $X$, which can be e.g. the energy
or the eccentricity, can be split into
a part corresponding to the effect from the PN forces including the GW reaction
and a part corresponding to additional effects caused by interactions with the
stellar environment:
\begin{equation}
\dv{X}{t} = \eval{\dv{X}{t}}_\text{PN} + \eval{\dv{X}{t}}_\text{env}.
\end{equation}
From this we find the cumulative effect of the environment as
\begin{equation}
\Delta_\text{env} X(t) = \int_{t_0}^t\eval{\dv{X}{t}}_\text{env} \dd{t}
= X(t) - \int_{t_0}^t\eval{\dv{X}{t}}_\text{PN} \dd{t}.
\end{equation}
This quantity is more robust than the instantaneous derivative 
$\eval{\dvi{X}{t}}_\text{env}$, as numerically differentiating the $X(t)$ derived
from a \ketju{} run to compute the instantaneous derivative amplifies any
numerical noise present in the data, which can easily hide the small effects of
the stellar environment.

To compute the effect of the PN forces
at a given time $t'$, i.e. $\eval{\dvi{X}{t}}_\text{PN}(t')$,
we perform a short integration of an
isolated binary with initial conditions taken from the \ketju{} computed
SMBH positions and velocities.
The integration is performed for about 30 orbits of the binary, and the
derivative at the middle of the integration interval is estimated by fitting a
line to the densely sampled evolution of $X$ over this period.
The initial conditions are chosen so that the middle of the integration interval
coincides with the desired time $t'$.

We evaluate the derivatives at about 50000 points spaced evenly in the semi-major
axis $a$ over the considered section of the \ketju{} results.
The cumulative effect $\int_{t_0}^t\eval{\dvi{X}{t}}_\text{PN} \dd{t}$ is then
found by integrating the instantaneous derivatives numerically.
With these choices this method of computing the instantaneous derivatives caused
by the PN terms appears to work well until the binary separation is under $\sim
100 R_s$, after which it fails as the evolution of the parameters starts to
become non-linear over the integration period.

\subsection{Semi-Analytic Isolated Evolution} \label{sec:isolated_evolutions}

For comparison against the SMBH binary evolution computed with \ketju{} we
consider two semi-analytic models used to describe binaries in simulations that
cannot resolve them.

The first model is the widely used \citet{1964PhRv..136.1224P} result,
which considers the secular evolution of the Keplerian orbital parameters 
of an isolated binary due to the leading radiation reaction term at 
\PN{2.5} level.
The binary is assumed to be otherwise Keplerian,
and the orbital parameters $a$ and $e$ are assumed to change only slowly.
The evolution of the semi-major axis $a$ is then given by
\begin{equation} \label{eq:peters_dadt}
\dv{a}{t} = - \frac{64}{5} \frac{G^3 m_1 m_2 M}{c^5 a^3 (1-e^2)^{7/2}}
                \biggl(1+\frac{73}{24} e^2 + \frac{37}{96}e^4 \biggr)
\end{equation}
and the eccentricity $e$ evolution by
\begin{equation} \label{eq:peters_dedt}
\dv{e}{t} = -\frac{304}{15}  \frac{G^3 m_1 m_2 M}{c^5 a^4
(1-e^2)^{5/2}} e \biggl( 1+\frac{121}{304} e^2 \biggr).
\end{equation}

\subsection{Semi-Analytic Scattering Model}
\label{sec:sa_scattering}

The main effect of the stellar environment on the evolution of the SMBH binary
is through the scattering of individual stars during close encounters.
These scattering events remove energy and angular momentum from the binary,
decreasing the semi-major axis $a$ and increasing the eccentricity $e$.
The second semi-analytic model considered here is a 
commonly used \citep[e.g.][]{2010ApJ...719..851S, 2017MNRAS.471.4508K} 
model describing this binary hardening process
\citep{1980AJ.....85.1281H, 1996NewA....1...35Q},
which gives the evolution of the semi-major axis and eccentricity as
\begin{equation} \label{eq:quinlan_a}
\dv{}{t}\biggl(\frac{1}{a}\biggr) = \frac{G\rho}{\sigma} H
\end{equation}
and
\begin{equation} \label{eq:quinlan_e}
\dv{e}{t} = -K a^{-1} \dv{a}{t} = K \frac{G\rho}{\sigma} H a,
\end{equation}
where $\rho$ is the stellar density and $\sigma$ the velocity dispersion.
Equation \eqref{eq:quinlan_a} can also be written in terms of the energy of the
binary as
\begin{equation} \label{eq:quinlan_dEdt}
\dv{E}{t} = -\frac{G M \mu}{2} \frac{G\rho}{\sigma} H,
\end{equation}
where $M=m_1+m_2$ is the total and $\mu = m_1 m_2/M$ the reduced mass of the
binary.
This form is more suited for our applications as the post-Newtonian corrections
modify the relation between the energy $E$ and semi-major axis $a$ compared to the non-relativistic Keplerian case.

The constants $H$ and $K$ are usually determined by three-body scattering
experiments \citep{1996NewA....1...35Q, 2006ApJ...651..392S}, but
\citet{2015MNRAS.454L..66S} find that similar values for the constants can also be recovered in N-body
simulations when the stellar density and velocity dispersion are evaluated at
the influence radius of the SMBH binary.
Traditionally, the influence radius $r_\text{inf}$ is defined by the condition that the
enclosed stellar mass is twice the binary mass, i.e. that
$M_*(r<r_\text{inf}) = 2 M_\bullet$.

\section{Gravitational Wave Computations}
\label{sec:gw_computation}
\subsection{The Gravitational Wave Background}
\label{sec:sa_gwb}

We focus our gravitational
wave computations on quantities relevant for observations with PTAs, which can
detect the low frequency GWs emitted by the massive systems studied in this
work.
The main observable quantity describing the stochastic GWB is the 
characteristic strain $h_c(f)$, which can be computed using either semi-analytic
or Monte Carlo (MC) methods when the distribution of sources and their GW emission is
known.

In the semi-analytic method, the characteristic strain $h_c(f)$ of the gravitational wave background at
frequency $f$ is given by 
\citep{2001astro.ph..8028P, 2007PThPh.117..241E}
\begin{equation}
h_c^2(f_o) = \frac{4G}{\pi c^2 f} \int \dd{z} \, n_c
\eval{\dv{E_{GW}}{f}}_{f=(1+z)f_o} ,
\end{equation}
where $n_c$ is the comoving number density of sources and the source
distribution is assumed to be spatially uniform.
As the source density $n_c$ is determined by the large-scale evolution of the
Universe, here we compute only $\dvi{E_{GW}}{f}$, the GW spectral energy density
(SED) of a source integrated over its lifetime, which is thus affected by any
processes speeding up or delaying the merger of the SMBH binaries.
In the MC methods of computing the characteristic strain 
\citep[e.g.][]{2008MNRAS.390..192S}, the characteristic strain is instead
computed by simply summing the instantaneous contributions from each source.
Thus all the environmental effects are included in the source distribution which
we do not consider here.

For computing the instantaneous GW flux and the SED of an SMBH binary 
we use two different methods depending on the regime:
in the weakly relativistic regime we save computational effort by 
applying analytically derived formulae to 
compute the GW spectrum from the orbital elements, 
while in the regime where the relativistic effects become significant we compute
the GW signal directly from the motion of the BHs.
This split between methods allows us to take advantage of the full \PN{3.5} accurate
dynamics used in \ketju{} while saving computational effort where possible
without sacrificing numerical accuracy.

\subsection{Semi-Analytic Spectrum}
\label{sec:sa_gw}

When the post-Newtonian effects are small, it is possible to compute the
contribution to the SED to good accuracy using analytic formulae derived
assuming a slowly varying Keplerian orbit.

A classic and widely used result derived by \citet{1963PhRv..131..435P} is the
power emitted in gravitational waves by a binary in the
lowest order quadrupole approximation.
For an eccentric orbit with eccentricity $e$ and orbital frequency $f_p$, the
GWs are emitted in all harmonics of the orbital frequency, with the $n$th
harmonic having the power
\begin{equation} \label{eq:peters_spectrum_luminosity}
P_n = \frac{32 G^{5/3} m_1^2 m_2^2 }{5 c^5 a^2} (2\pi f_p)^{2} g(n,e).
\end{equation}
The function $g(n,e)$ is given by
\begin{equation}
\begin{aligned}
g(n,e) = \frac{n^4}{32} \bigg\{&
\Big[ J_{n-2}(ne) - 2e J_{n-1}(ne)\\ 
    &+ \frac{2}{n} J_n (ne) + 2 e J_{n+1}(ne)- J_{n+2}(ne)
\Big]^2\\
&+(1-e^2)\left[ J_{n-2}(ne) - 2 J_n(ne) + J_{n+2}(ne)\right]^2\\
&+ \frac{4}{3 n^2} J_n(ne)^2
\bigg\},
\end{aligned}
\end{equation}
where $J_n$ is the Bessel function of the first kind of order $n$.
Since in this regime the PN effects are small, we may simply use the PN
quasi-Keplerian parameters in place of the Keplerian parameters, thus taking $f_p = f_r$.

From the instantaneous GW power and the known evolution of the orbital elements
the SED can be computed as \citep{2007PThPh.117..241E}
\begin{equation}
\dv{E_{GW}}{f} = \sum_{n=1}^\infty \left.
\frac{P_n(f_p, e)\tau_{GW}(f_p,e)}{n f_p}  \right|_{f_p=f/n},
\end{equation}
where
\begin{equation} \label{eq:tau_GW_def}
\tau_{GW} = f_p \dv{t}{f_p} = \dv{t}{(\ln{f_p})}
\end{equation}
is the gravitational wave timescale.
Implementing this formula for computations from numerical simulation outputs is
mostly straightforward: for each frequency $f$ we truncate the sum at some
sufficiently high number of harmonics,
and interpolate the eccentricity $e$ and
gravitational wave timescale $\tau_{GW}$ to the required orbital frequencies
$f_p$.
Here $e$ and $\tau_{GW}$ are treated as functions of the orbital frequency
$f_p$ instead of time $t$, which is possible as generally $f_p(t)$ is
monotonic in the relevant phase of the binary evolution.
The number of harmonics used for the computation is taken to be 750 in this
work, which gives converged results for the eccentricities encountered. 

The computation of $\tau_{GW}$ from the numerical data requires some
care, as naively numerically differentiating the time-frequency data
will amplify any noise caused for example by interactions with the environment
to unusable levels.
To avoid this, we instead compute the timescale by integrating the time spent in
logarithmic bins of orbital frequency.
If two adjacent data points belong to the same bin the whole interval is
assigned to that bin, while for points in different bins the time interval is
divided over the spanned frequency interval.
We find that using about 50 bins per decade in frequency gives results that
match well with the differentiation of smooth data.

The error in applying these leading order or \PN{0} results is expected to be of
order \PN{1}, and in \autoref{sec:sa_method_accuracy} 
we find that the error between this method and the more accurate
method described in the next subsection is of the order of a few percent when
the PN expansion parameter $\epsPN \approx 0.01$, in line with expectations.

Similar methods are commonly extended also beyond the regime where PN effects are
minor all the way down to the
final orbits before the BHs merge 
\citep[e.g.][]{2009ApJ...695..455B, 2017MNRAS.471.4508K}.
This may work well if the orbital evolution is consistently performed in the
same approximation, but coupled with \PN{3.5} accurate orbits this leads to clearly
erroneous results.
There are also analytic results for PN accurate spectra available in the
literature \citep[e.g.][]{2010PhRvD..82l4064T,2018PhRvD..98j4043K}, however these
are quite cumbersome to apply, and we find the methods of the next subsection
to be more convenient.

\begin{figure*}
\plottwo{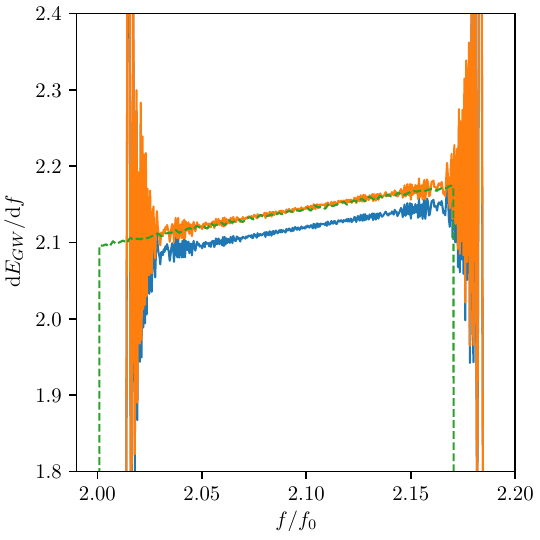}{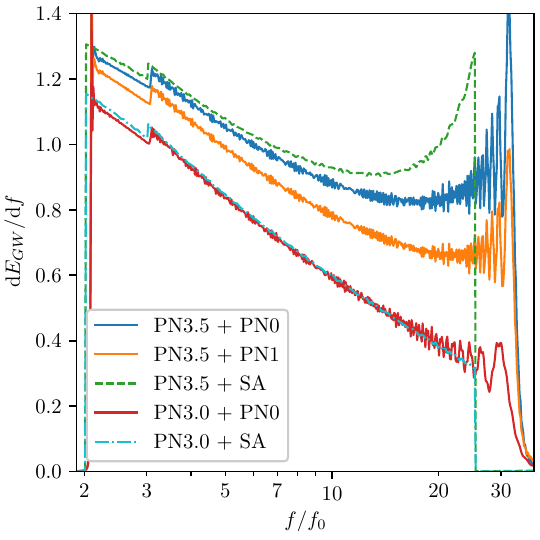}
\caption{
The emitted GW energy spectrum (in code units)
computed for \PN{3.5} accurate binary motion using the semi-analytic method
(\PN{3.5} + SA, dashed green line)  and directly from the waveform using either only the leading
quadrupole terms (\PN{3.5} + \PN{0}, solid blue line) or including also \PN{1} corrections to the
waveform (\PN{3.5} + \PN{1}, solid orange line).
Only the smoothed result is shown for the direct waveform calculations, but some
artefacts caused by the finite signal length are still visible near the edges.
The left panel shows the main $n=2$ harmonic of the signal from a SMBH binary with
an initial semi-major axis of $a=250 R_s$ and eccentricity $e=0.2$ computed over
\num{2e4} orbits.
The right panel shows the full spectrum from a binary with initial $a= 40 R_s$
and $e=0.1$ evolved until $a \approx 6 R_s$.
The frequency is given in units of the initial radial frequency $f_0 =
f_r(t=0)$.
The right panel also shows in addition the smoothed quadrupole spectrum computed from the
waveform
(\PN{3.0} + \PN{0}, solid red line) and using the semi-analytic method (\PN{3.0} + SA, dot-dashed cyan line)
from an otherwise identical
binary evolution but using only PN-corrections up to \PN{3} level.
Note the logarithmic frequency scale in the right panel.
}
\label{fig:dEdf_comparison}
\end{figure*}

\subsection{Direct Waveform Calculation}
\label{sec:fft_gw}

When the effects of the PN-corrections on the now non-Keplerian binary orbit become significant, 
the methods of the previous subsection are no longer reliable.
In this regime we instead directly compute the emitted waveform, and use a
Fourier transform to compute the spectrum.

In principle this method could be used over the entire binary evolution, but
for most part the semi-analytic method is sufficiently accurate.
Further, if the binary eccentricity is high,
the timestep required to resolve the
highest harmonics of the orbital frequency becomes very small resulting in considerable computational cost.
For these reasons we limit this method to the short period before the BH binary
merger where it is necessary for accurate results.
In this regime the effect of the environment is also negligible, allowing us to
easily compute a densely outputted evolution of the SMBH binary in isolation using the
\ketju{} integrator.
Inclusion of the environment at this stage would also be possible, but this would
require suitable schemes to resample the stellar population to lower mass
particles in order to avoid artefacts in the spectrum caused by spurious 
unphysically strong interactions from massive stellar particles. 

In computing the GW waveform emitted by the SMBH binary we follow 
\citet{poisson2014gravity}.
The leading contribution to the strain tensor $h_{ij}$ is the quadrupole term
\begin{equation} \label{eq:quadrupole_gw}
h_{ij} = 
4 \frac{G m_1 m_2}{c^4 d M} \Bigl(V_i V_j - G M \frac{R_i R_j}{R^3} \Bigr),
\end{equation}
where the relative velocity $V_i$ and separation $R_i$ are computed directly
from the numerical integration of SMBH trajectories.
The observer distance $d$ only scales the result, and can in practice be ignored
in these computations.
The radiation reaction force corresponding to this part appears as the
$\vc{a}_{\PN{2.5}}$ term in the equations of motion, and to maintain consistency with
the \PN{3.5} accurate equations of motion we also include PN-corrections to the
waveform up to \PN{1} order beyond the leading \PN{0} contribution of equation
\eqref{eq:quadrupole_gw}.
The formulae for these corrections for a binary system are lengthy, and are not
reproduced here but can be found in \citet{poisson2014gravity}.
Note that by convention the waveform PN orders are offset from the corresponding
effects in the equations of motion by $2.5$ orders.

The strain tensor $h$ obtained as a function of  time is then projected into the
physically observed polarizations $h_+$ and $h_\times$, which are related to
emitted GW power as 
\begin{equation} \label{eq:GW_power_waveform}
\dv{E_{GW}}{t} = 
\frac{d^2 c^3}{16 \pi G} \int \bigl(
    (\partial_\tau h_+)^2 +  (\partial_\tau h_\times)^2\bigr) \dd{\Omega},
\end{equation}
where $\tau = t - d/c$ is the retarded time, which can here  be replaced with
the coordinate time $t$ due to the arbitrary observer distance.
In terms of the Fourier coefficients $\tilde{h}_{+,\times}(f)$ the time
dependent signals can be written as 
\begin{equation}
h_{+,\times}(t) =
\sum_k \sqrt{2} \abs{\tilde{h}_{+,\times}(f_k)} \sin(2\pi f_k t + \phi_k),
\end{equation}
so the total energy emitted over a time period of length $T$ can be written as
\begin{equation}
\begin{aligned}
\Delta E_{GW} &= \int_{t_0}^{t_0+T} \dd{t} \dv{E_{GW}}{t} \\
&= \frac{d^2 c^3 T}{16\pi G} \sum_{\substack{k\\p=+,\times}} (2\pi f_k)^2 
\int \abs{\tilde{h}_p (f_k)}^2 \dd{\Omega}.
\end{aligned}
\end{equation}
As this can also be written as 
\begin{equation}
\Delta E_{GW} = \sum_k  \frac{\Delta E_{GW}(f_k)}{\Delta f} \Delta f,
\end{equation}
the SED at the frequency $f_k$ is given by the corresponding term in the
sum
\begin{equation}
\begin{aligned}
\dv{E_{GW}(f_k)}{f} \approx & \frac{\Delta E_{GW}(f_k)}{\Delta f} \\
=& \frac{c^3(2 \pi f_k T d)^2}{16 G}  
\sum_{p=+,\times} \int \abs{\tilde{h}_p (f_k)}^2 \dd{\Omega},
\end{aligned}
\end{equation}
where $\Delta f = f_{k+1} -f_{k}  = 1/T$ is the frequency resolution of a
discrete Fourier transform of a signal with evenly spaced samples.

To numerically compute these quantities we use the fast Fourier transform (FFT)
algorithm, which allows computing the coefficients
$\tilde{h}_{+,\times}$ efficiently from evenly spaced samples of the waveform.
The integral over the solid angle $\Omega$ is conveniently performed
numerically, although at least when ignoring the direction dependent PN
corrections to the waveform it can also be treated analytically.
The changing frequency of the signal leads to rapid oscillations in the FFT
results, which we suppress by averaging them over a small window.
Additionally, the abrupt changes at the ends of the signal lead to some
artefacts.
These could be reduced by windowing, but this would also reduce the accuracy of
the computed energy of the GW signal.
From the Fourier transformed signals we may also evaluate the
instantaneous power from equation \eqref{eq:GW_power_waveform},
using the relation between derivatives and Fourier transforms.
Evaluating the time derivatives via Fourier transforms is also more accurate for
periodic signals than using finite difference techniques.

\subsection{Accuracy of the Semi-Analytic Method}
\label{sec:sa_method_accuracy}

In order to confirm that the semi-analytic method gives good results in the mildly
relativistic regime and to find the limits of its applicability
we compare it to the direct waveform calculation over short sections of the
evolution of an isolated binary.
This allows also evaluating the effect of including the additional PN
corrections to the waveform instead of simply using the quadrupole formula
\eqref{eq:quadrupole_gw}.

For the mildly relativistic regime we compute the orbit of an isolated SMBH binary with a mass
ratio of $q=5:1$
and initial semi-major axis $a = 250 R_s$ and eccentricity $e = 0.2$ over
\num{2e4} orbital periods.
This fairly large number of orbits allows the orbital frequency to evolve enough
for the GW spectrum to be resolved instead of effectively 
consisting of $\delta$-spikes which are problematic numerically.
The post-Newtonian expansion parameter is approximately $\epsPN \approx 1/200$
in this case, so the effect of the additional PN effects not taken into account
in the semi-analytic model can be expected to be of the order of half a percent.

The left panel of \autoref{fig:dEdf_comparison} shows the resulting spectrum
around the strongest $n=2$ harmonic.
Apart from the small shift in frequency, which is of the order of $\epsPN$,
the semi-analytic result matches well with the direct waveform computation using
the full \PN{1} accurate waveform.
This agreement with the \PN{1} waveform may seem surprising, as the semi-analytic
method only includes the \PN{0} quadrupole radiation, but this can be simply
explained based on energy conservation.
In this regime the binary is still close to Newtonian and 
the main effect of the \PN{1} waveform corrections is to
maintain energy balance with the \PN{3.5} radiation reaction terms.
As a result, the semi-analytic method,
which is also constructed around energy conservation of a Newtonian binary,
gives matching results.
The frequency shift is caused by the precession of the binary, which leads to
the GW frequencies differing from integer multiples of the radial frequency used
for the calculation.
While this shift could in principle be corrected for, we find that it is insignificant when
the spectrum is computed over the entire binary evolution.

To evaluate the methods in the highly relativistic regime we compute the
evolution of the binary from an initial semi-major axis of $a = 40 R_s$ down to $a
= 6 R_s$, where the PN equations of motion become unreliable.
The initial eccentricity is set to $e \approx 0.1$, which is comparable to
the eccentricities found in our \ketju{} runs at this binary separation.
The resulting spectra in the right panel of \autoref{fig:dEdf_comparison}
clearly show the increasing significance of consistently computing the waveform
from the actual binary motion as well as including the PN-corrections to the waveform,
with the differences between the methods growing to tens of percents towards the
end of the simulation.

The error of the semi-analytic method compared to the full PN waveform 
is about 5\% at the point where all the major harmonics are visible in the
spectrum, but somewhat lower when compared to the \PN{0} waveform.
In this regime it is also of interest to look at the significance of the higher
order radiation reaction terms at \PN{3.5} level.
The right panel of \autoref{fig:dEdf_comparison} includes also the quadrupole
spectrum from an otherwise identical binary evolved using PN-corrections only
up to \PN{3} level.
The amplitude of the spectrum is clearly different from the full \PN{3.5} result,
showing the importance of including the higher order radiation reaction term.
Interestingly, the semi-analytic method yields essentially exact results
compared to the \PN{0} waveform in this case.
This indicates that the differences between the semi-analytic method and the
direct waveform methods are at least partly due to the \PN{3.5} radiation reaction term
causing significant changes in the motion of the binary.

\begin{figure*}
\plotone{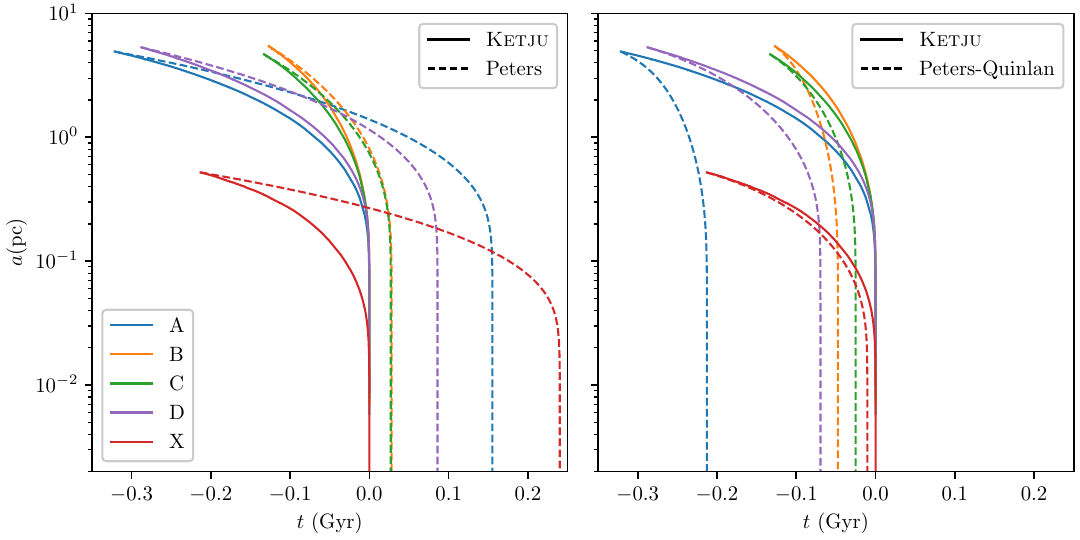}
\caption{
Evolution of the semi-major axis $a$ as a function of time.
The solid lines show the \ketju{} runs 
while the dashed lines show the Peters (left panel) 
and Peters-Quinlan (right panel) semi-analytic comparison models.
The times have been shifted so that the merger of the BHs in the \ketju{} runs
takes place at $t=0$, corresponding to a shift of \SIrange{1}{2.5}{Gyr}.
}
\label{fig:a_evolution}
\end{figure*}

\begin{figure*}
\plottwo{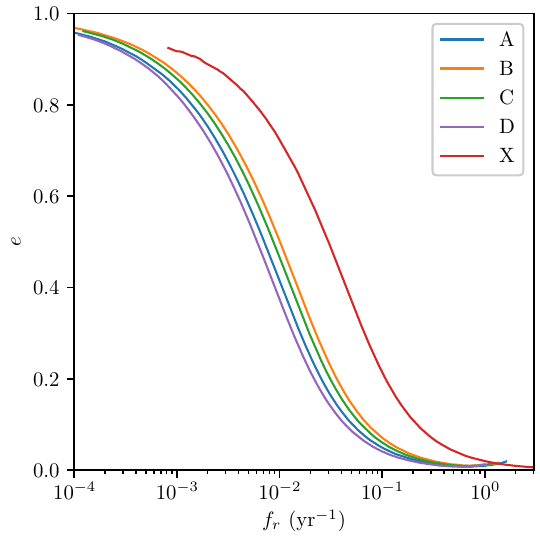}{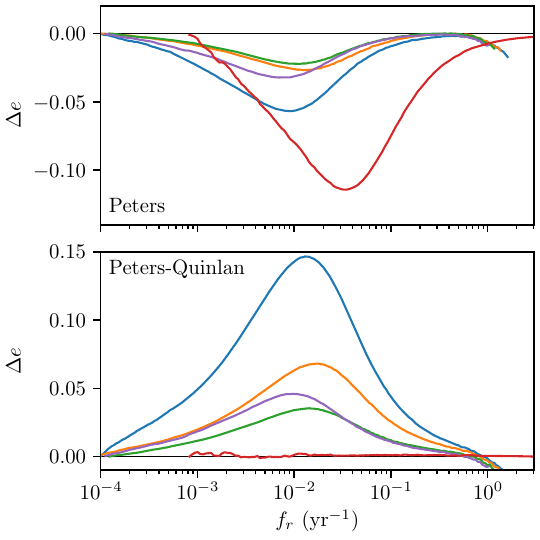}
\caption{
Eccentricity evolution as function of radial orbital frequency $f_r$.
The left panel shows the results from the \ketju{} runs, and the right panel shows the 
difference to the \ketju{} runs for the Peters model (top right) and the
Peters-Quinlan model (bottom right). 
}
\label{fig:e_freq_plots}
\end{figure*}

\section{Results}
\label{sec:results}

\subsection{Orbital Evolution}
We begin by studying the evolution of the orbital elements of the SMBH binaries in the
\ketju{} runs and compare them to the parameters predicted by
semi-analytic models describing both isolated binaries and
binaries interacting with a population of stars.
The semi-analytic comparison models are started from the initial conditions at the start of the
analyzed section, i.e. using the values of $a_0$ and $e_0$ listed in
\autoref{table:runs}.
This is similar to what would be done in a standard softened simulation when the
SMBH binary separation has decreased to the order of a gravitational softening length.

Isolated binaries are modeled using the model described in 
\autoref{sec:isolated_evolutions}, hereafter the \emph{Peters} model,
which describes a Keplerian
binary perturbed by the leading gravitational radiation reaction term.

The interaction with stars is included by combining the Peters model with the scattering 
model of \autoref{sec:sa_scattering}, to produce what we call the 
\emph{Peters-Quinlan} model.
This model requires as inputs the
stellar density and velocity dispersion, which are computed from the
\ketju{} run results in a sphere within the influence radius of the binary.
For these calculations we use the values of $H$ and $K$ parameters determined by the fitting
formulae of \citet{2006ApJ...651..392S}, which allows us to see how much the
behavior of the \ketju{} runs differs for typically used literature values.
The values of $H$ are computed using the fit for a circular binary,
as \citet{2006ApJ...651..392S} do not tabulate the coefficients for the $H$ fit
for eccentric binaries.
Their results however indicate that at higher eccentricities the value of $H$ should tend to increase slightly.

The time evolution of the semi-major axis of the orbits in
the \ketju{} runs and the semi-analytic comparison models is shown
in \autoref{fig:a_evolution}.
The evolution of the binary is mainly driven by GW emission at this
stage, but the environment still has a non-negligible effect.
This can be seen by studying the isolated Peters models, in which the SMBHs merge
significantly later than in the \ketju{} runs.
On the other hand, the Peters-Quinlan models merge significantly earlier than the
\ketju{} runs, which is caused by the fact that the literature values of the 
$H$ and $K$ parameters from \citet{2006ApJ...651..392S} significantly overestimates
the effects of stellar scattering for our \ketju{} runs with depleted stellar cores (runs A--D).
This is most obvious in the case of run A, while in the case of run X the match
between the \ketju{} run and the Peters-Quinlan model is reasonably good.

The eccentricity evolution of the runs is shown in \autoref{fig:e_freq_plots} as
a function of the orbital frequency $f_r \approx \sqrt{GM/a^3}/(2 \pi)$, as
this allows a comparison between the models with different merger times and
also relates the eccentricity to the frequency of emitted GWs.
Runs A--D all show fairly similar behavior with high eccentricities in the
range $e \approx 0.96 \text{--} 0.98$ at the initial semi-major axis $a \sim \SI{5}{pc}$
corresponding to an orbital frequency of $f_r \sim \SI{e-4}{yr^{-1}}$
where we begin considering these runs.
Run X has a slightly lower eccentricity at its initial $a \approx \SI{0.5}{pc}$,
but it too is over $0.9$.

The differences between the two semi-analytic models reveal an additional reason for the
different merger times, in addition to the direct decay of the semi-major axis
due to stellar scattering: the eccentricities of the Peters model are considerably lower
and those of the Peters-Quinlan model higher than in the \ketju{} runs.
As can be seen from equation \eqref{eq:peters_dadt}, the dependence of the
evolution on the eccentricity is highly non-linear, and especially at high
eccentricities even small differences lead to large effects in the merger time.
Notably the difference in the evolution of the eccentricity is smallest in the
case of the Peters-Quinlan model run X and largest in the case of run A, which
show also correspondingly the smallest and largest difference in the SMBH 
merger times, respectively.

\subsection{Strength of Environmental Effects}

\begin{figure*}
\plottwo{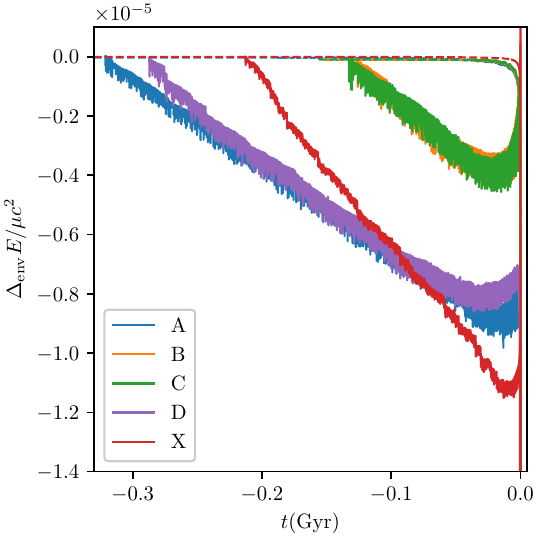}{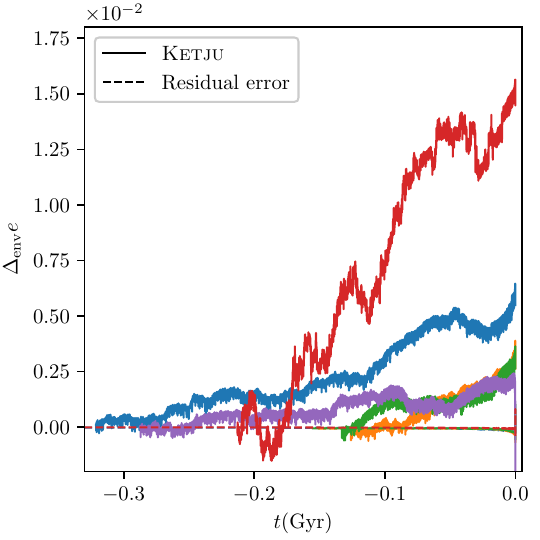}
\caption{
The evolution of the energy per unit reduced mass $E/\mu c^2$ (left) and the eccentricity $e$
(right) of the SMBH binary due to the interaction with the stellar environment.
The contribution from the PN terms has been removed as explained in section
\ref{sec:extracting_env}, with the dashed lines showing the residual from the
subtraction process for an isolated evolution.
The removal process works well until the final stages of the merger where the
difference in energy starts to grow rapidly.
There is a remaining secular evolution in the \ketju{} energy which appears
to be consistent with the simple semi-analytic scattering models. 
}
\label{fig:cumulative_env_effect}
\end{figure*}

\begin{deluxetable}{cccccc}
\tablecaption{Fitted and literature values of the hardening parameters
\label{table:H_K_fits}
}
\tablehead{
\colhead{Run} 
& \colhead{$H_\ketju{}$} & \colhead{$K_\ketju{}$} 
& \colhead{$H_S$} & \colhead{$K_S$} 
& \colhead{$G \rho/\sigma$} \\
&&&&&\colhead{ ($\SI{e-11}{pc^{-1} . yr^{-1}}$)}
}
\startdata
A   & 2.18 & 0.032   &  16.2 & 0.056  & 5.49 \\
B   & 4.38 & 0.049   &  16.3 & 0.059  & 2.92 \\
C   & 5.98 & 0.052   &  16.4 & 0.068  & 1.68 \\
D   & 8.44 & 0.027   &  16.6 & 0.068  & 1.03 \\
X   & 11.5 & 0.098   &  14.4 & 0.054  & 24.4 \\
\enddata
\tablecomments{
The values of the hardening parameters $H_\ketju{}$ and $K_\ketju{}$ 
of \autoref{sec:sa_scattering} have been computed by linear fits to the \ketju{}
data with the effects of the \PN{3.5} level corrections subtracted (see
\autoref{sec:extracting_env}).
The values used for the Peters-Quinlan comparison models ($H_S$, $K_S$)
computed from the fitting formulae of \citet{2006ApJ...651..392S} are also
listed, as well as the value of $G\rho/\sigma$ measured from the simulations
at the influence radii of the SMBH binaries. 
}
\end{deluxetable}

To quantify the strength of the environmental effects, we use the method of 
\autoref{sec:extracting_env} to subtract the dominant effects of the
PN-corrections and recover the residual effect caused directly by interactions
with the stellar environment.
The results for the energy and eccentricity of the SMBH binaries are shown in 
\autoref{fig:cumulative_env_effect}.
The energies show a clear linear evolution with time, in agreement with the
semi-analytic model equation \eqref{eq:quinlan_dEdt}.
The eccentricities also show a secular evolution with the correct sign, but the
effect is fairly weak.

We perform linear fits to the data in \autoref{fig:cumulative_env_effect} to determine the 
values of the $H$ and $K$ parameters of the scattering model that would
correctly describe the binary evolution when the stellar densities and velocity
dispersions are derived from the \ketju{} runs.
Note that while equation \eqref{eq:quinlan_dEdt} implies that the change in energy
is linear in time, equation \eqref{eq:quinlan_e} implies that the change is
eccentricity is linear in the integrated semi-major axis:
\begin{equation}
    \Delta_\text{env} e(t) = K H \frac{G \rho}{\sigma} \int_{t_0}^t a \dd{t}.
\end{equation}

The results are given in \autoref{table:H_K_fits}, which also shows the
$H_S$- and $K_S$-values used for the Peters-Quinlan comparison models, derived from 
\citet{2006ApJ...651..392S}.
In the case of run X the fitted value of $H$ is close to the value determined
from the \citet{2006ApJ...651..392S} fitting formulae, being about 25\% lower.
This is comparable to the results of \citet{2015MNRAS.454L..66S}, who found that
N-body simulations tend to produce approximately 30\% lower values of $H$ than
those found from idealized 3-body scattering experiments, which were used by 
\citet{2006ApJ...651..392S}.

For runs A--D the agreement between the fitted $H$ parameter and the one used
for the Peters-Quinlan model is much poorer, with up to a factor of eight
difference in the case of run A. This is not entirely unexpected as these runs describe 
massive early-type galaxies with very large low density stellar cores.
Thus, the derived small values of $H$ are caused by the fairly empty loss cones in these
runs, as most of the stars on radial trajectories have already been ejected at this
stage by the strong core scouring effects of the SMBH binary \citep{2019ApJ...872L..17R}. 

The fitted values of $K$ are slightly lower for runs A--D, as can be expected due to
the low density cores.
However, the differences are not quite as large as in the case of the $H$ parameter.
On the other hand, the fitted value of $K$ for run X is almost twice as large as
the literature value used for comparison.
This is likely partly due to effect of eccentricity on the parameters,
as the comparison value was computed for a circular binary.
For run X the eccentricity evolution between the Peters-Quinlan model and the \ketju{} run
agrees very well, despite slight differences in the scattering parameters (see Table \ref{table:H_K_fits}).
This is made possible since there is a certain degree of degeneracy between the $H$ and $K$ parameters in the 
Peters-Quinlan model, allowing very similar evolutions to be produced for a range of parameter values.

We also computed comparison Peters-Quinlan models using the parameter values fitted
from the \ketju{} runs.
For all runs the results are accurate at a similar level as the run X comparison model using
literature values discussed above, i.e. at the level of a few percent.
As the behavior is so similar, we opt to not display the results of these calculations.

\subsection{Gravitational Wave SEDs}

\begin{figure*}
\plottwo{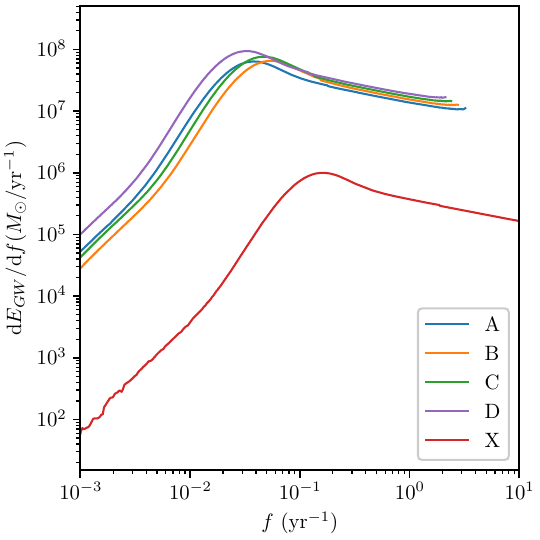}{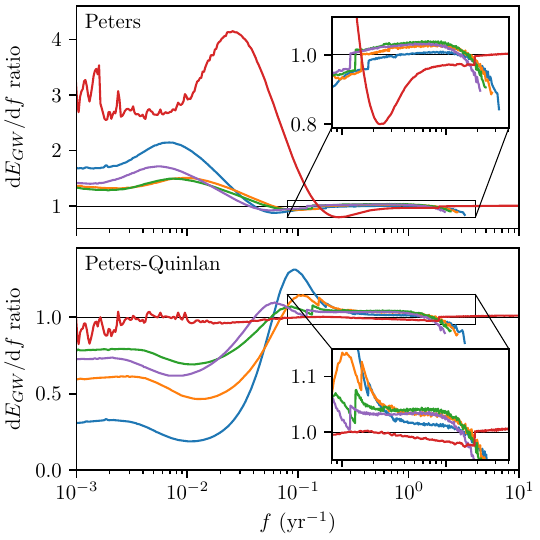}
\caption{
Total emitted GW energy per unit frequency, i.e. the GW SED,
for the \ketju{} runs (left),
and the ratios of the semi-analytic Peters model (top right)
and the semi-analytic Peters-Quinlan model (bottom right) results to the \ketju{}
result.
The zoom-ins highlight the frequency range most relevant for PTA observations
$(f\gtrsim \SI{e-1}{yr^{-1}})$. 
}
\label{fig:dEdf}
\end{figure*}

\begin{figure*}
\plottwo{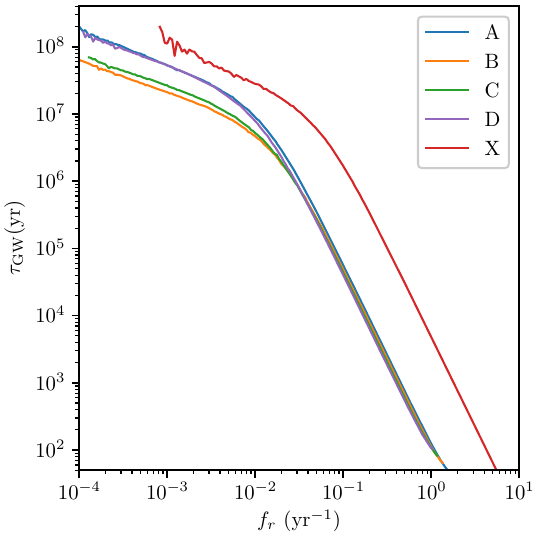}{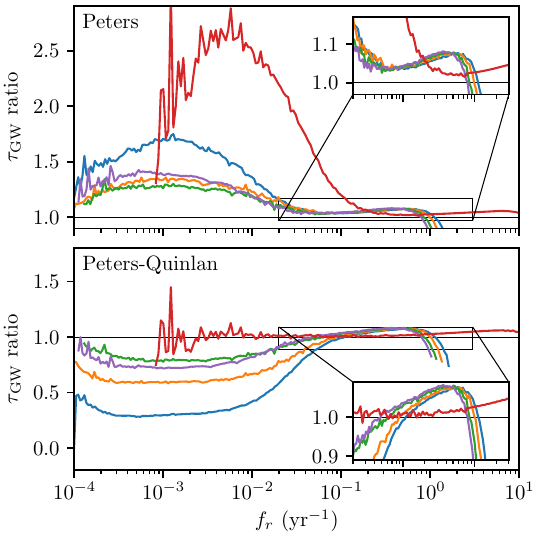}
\caption{
The gravitational wave timescale $\tau_{GW} = \dv{t}{\ln{f_r}}$ of the
\ketju{} runs (left),
and the ratios of the semi-analytic Peters model (top right)
and the semi-analytic Peters-Quinlan model (bottom right) results to the \ketju{}
result.
The zoom-ins highlight
the frequency range most relevant for PTA observations $(f\gtrsim \SI{e-1}{yr^{-1}})$. 
}
\label{fig:tauGW}
\end{figure*}

Next we focus on the GW signal emitted by the SMBH binaries, and the differences
between the signal computed from the \ketju{} runs and the semi-analytic comparison
models.
In computing the GW spectrum from the \ketju{} runs we use the semi-analytic
method of \autoref{sec:sa_gw} until the SMBH binary semi-major axis is
$\sim 100 R_s$, after which we
switch to the direct waveform method of \autoref{sec:fft_gw}
and compute the isolated evolution of the binary for the final $a \lesssim 100 R_s$.
The exact point where the switch happens depends on the eccentricity of the
binary, as all the major harmonics of the initial orbital frequency need to be
visible in the spectrum computed from the waveform.
Based on \autoref{sec:sa_method_accuracy}, 
the error of the semi-analytic GW method should be of the order of a
few percent at the switchover point, as the error scales with $\epsPN \propto
a^{-1}$.
For computing the GW signal from the semi-analytic Peters and Peters-Quinlan
models, we naturally utilize the semi-analytic method throughout.

The energy spectrum of gravitational waves as emitted over the lifetime of the
binary is shown in \autoref{fig:dEdf}.
The spectra are all fairly similar due to the similar orbital evolution, and
show the characteristic peaked shape of an initially highly eccentric binary.
The differences between the runs are mainly explained by the different total
black hole masses and mass ratios, as well as the different eccentricities.

At high frequencies, where the binaries have mostly circularized $(e\sim 0)$, the spectra
are fairly similar.
For runs A--D the full PN FFT computation of the spectrum gives results that differ by only a 
few percent from the results computed using the Peters evolution combined with the
semi-analytic GW method until just before the
highest modeled frequencies when the binary is at a separation of $R=6 R_s$
where the \ketju{} simulation is stopped.
The Peters-Quinlan model gives differences of a similar magnitude for these runs,
which is expected as in this phase these massive binaries are highly relativistic
and the differences in the spectrum are mainly due to PN effects.
The effect of the environment in the spectrum is larger for run X, where the
Peters model shows a difference of almost 20\% at $f=\SI{0.2}{yr^{-1}}$,
while the Peters-Quinlan model agrees at the percent level.
This is mainly due to the larger eccentricity of the binary in this frequency range,
and in particular the difference in eccentricity in the Peters model, which shifts the peak of the spectrum.
The agreement at slightly higher frequencies where the binary has circularized is
much better, with the semi-analytic models matching almost exactly the direct
waveform computation results.
At the point when the GW calculation is switched to the direct waveform method, 
the differences to the semi-analytic GW method are of the order of few percent, consistent with
the PN expansion parameter being $\epsPN \sim 0.01$ at that point.
This leads to the small jump that can be seen in the difference to the
semi-analytic comparison models (see the zoom-ins in Figures \ref{fig:dEdf} and \ref{fig:tauGW}).

At lower frequencies where only the semi-analytic GW computations are performed on the 
\ketju{} simulations, the isolated Peters models lead to larger amplitudes with
peaks at lower frequencies.
The resulting relative differences are mostly around $50\%$, but can reach values of 
up to 300\% in the case of run X.
These large differences are due to the faster hardening of the binary due to the
stellar environment, which is not accounted for in the Peters models and 
in fact the Peters-Quinlan models which include this effect are
in much better agreement with the \ketju{} results.
The most significant differences between the \ketju{} and Peters-Quinlan results
occur in the case of run A, where the Peters-Quinlan result is about 70\% lower.
In this case the hardening parameter used for the Peters-Quinlan model was 
already found to be 5 to 8 times
larger than the effective ones determined from the \ketju{} run, so the large
difference is not surprising.
The other runs (B--D) with large differences in the hardening parameters show 
similar behavior, with differences of around $40\%$.
In any case the differences occur mainly at relatively low GW frequencies
that are unobservable for the foreseeable future.

The differences in the GW spectrum are mainly due to differences in the
gravitational wave timescale $\tau_{GW}$, shown in \autoref{fig:tauGW},
and also in part due to the differences in orbital eccentricity.
The evolution of eccentricity determines how the emitted GWs are spread out over
the spectrum, which also determines the location of the peak of the spectrum.
Here too the effects of the environment are obvious, with the Peters models
having longer and the Peters-Quinlan models shorter GW timescales than the \ketju{}
runs.
The differences are mostly between 10 and 60 percent at orbital
frequencies below $\SI{0.1}{yr^{-1}}$, with run X showing both the highest
difference of $\sim 150\%$ to the Peters model and almost no difference to the
Peters-Quinlan model.
At high frequencies both of the semi-analytic Peters formulae based comparison models show
slightly longer timescales, which is due to the inclusion of higher order PN
effects in \ketju{}.

\subsection{Gravitational waves and energy conservation}

\begin{figure}
\plotone{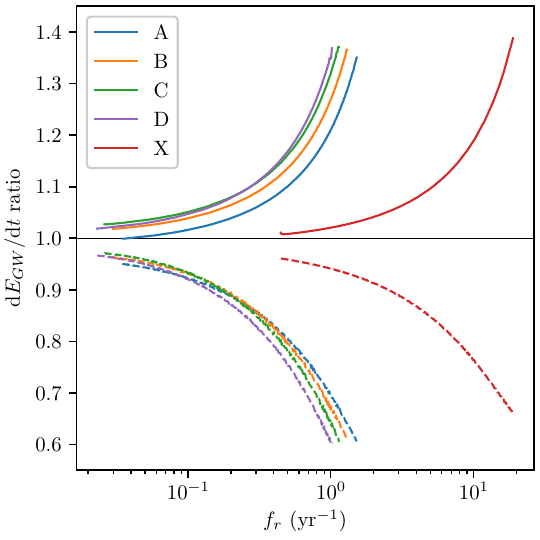}
\caption{
The ratio of the instantaneous GW fluxes computed using less 
accurate approximations to the \PN{3.5} motion + \PN{1} waveform used
for the final part of the GW SED calculations.
The solid lines show the result of the \PN{3.5} motion + \PN{0} (quadrupole) 
waveform calculation, while the dashed lines show the result of the 
semi-analytic Peters formulae calculation.
The differences between the runs are due to the different mass ratios and
slightly different eccentricities.
}
\label{fig:GW_flux}
\end{figure}

The good agreement at the percent level
between the spectra computed using the Peters formulae and
the full PN computation at high frequencies may seem surprising, since the
strongest effects of the PN-corrections should appear there.
If we look at the instantaneous GW flux, shown in \autoref{fig:GW_flux},
the difference between the methods does indeed increase with increasing
frequency, reaching tens of percent.
However, this is countered by the slightly slower evolution of the orbit
according to the Peters formula, which can be seen as a slightly longer
timescale at high frequencies in $\tau_{GW}$ in \autoref{fig:tauGW}.
Fundamentally, the similarity of the spectra computed with the different methods
for an isolated binary follows from the conservation of energy, which is
exact by construction in the Peters evolution and approximate in the \PN{3.5} evolution,
as was also noted in \autoref{sec:sa_method_accuracy}.
\autoref{fig:GW_flux} also shows the importance of including the waveform PN1
corrections as well, as otherwise the flux can be overestimated by over 10\%.

\begin{figure}
\plotone{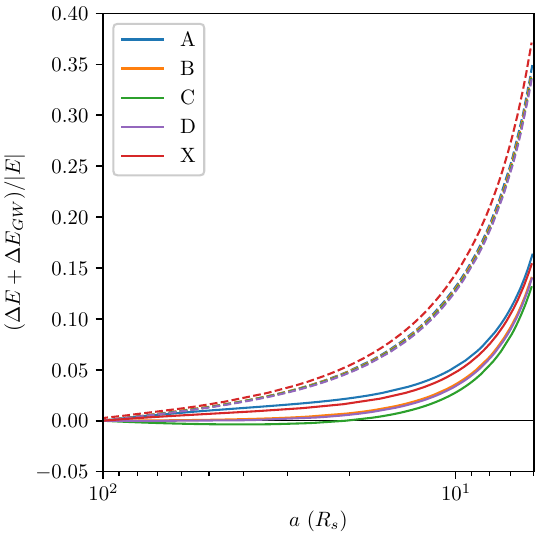}
\caption{
The relative error in total energy conservation during the final
isolated evolution of the binaries
when the emitted GW energy is computed from the \PN{1} waveform (solid)
or the \PN{0} quadrupole waveform (dashed).
}
\label{fig:energy_conservation}
\end{figure}

Another point relating to the energy conservation of the system
when using PN dynamics to compute the binary evolution and GW
emission in the final stages of an SMBH coalescence is the fact that the energy of the
system is conserved only approximately.
This limits the applicability of the method, as large violations of energy 
conservation clearly signify incorrect results.
\autoref{fig:energy_conservation} shows the total relative error in the energy
conservation of the system during the final stages of the
merger.
The orbital energy is computed using the PN accurate energy of equation
\eqref{eq:pn_energy_ang_momentum}, while the emitted GW energy is computed by 
integrating the instantaneous energy flux computed from the waveform.
For most part the full computation including the PN waveform corrections gives
an error of under 5\%, and only in the
very final stages does the error begin to rapidly increase.
This suggest that the results computed using this method are fairly
reliable down to around $a \sim 10 R_s$ or to orbital frequencies of
$f_r \sim (10^{10} M_\sun/M) \,\si{yr^{-1}}$, which also corresponds to the part
in the spectrum in \autoref{fig:dEdf} where the direct waveform and the
Peters results begin to again diverge.
When the GW emission is computed using only the quadrupole formula the error is
naturally greater.
A similar error in energy conservation for the full PN computation 
has also been observed by \citet{2012CQGra..29x5002C}.

\section{Discussion}
\label{sec:discussion}

\subsection{SMBH eccentricity Distribution}

Since the eccentricity affects strongly both the 
merging time of the SMBHs \citep{1964PhRv..136.1224P} and the emitted GW spectrum \citep{2007PThPh.117..241E}, 
understanding the distribution of eccentricities of SMBH binaries is essential for correctly 
predicting the emitted GWB. 
As can be seen from Table \ref{table:runs} all the \ketju{} runs studied in this paper show 
high initial eccentricities in excess of $e>0.9$ at the point where GW emission and 
other relativistic effects start to become important. The encounter orbits of our galaxies 
are nearly parabolic as motivated by cosmological simulations \citep{Khochfar2006} and high initial 
eccentricities have also been found in earlier studies employing \ketju{} \citep{2017ApJ...840...53R,2018ApJ...864..113R}.  

Previous works studying the evolution of SMBH binaries in a galaxy merger
using a smaller number of particles 
have also in general found high binary eccentricities, $e\gtrsim 0.8$ \citep[e.g.][]{2009ApJ...695..455B, 2011ApJ...732L..26P, 2011ApJ...732...89K,2018A&A...615A..71K}. The initial 
binary eccentricity is affected by the initial condition setup, with strongly radial merger orbits 
producing in general higher eccentricity binaries. Another factor that could affect the eccentricity distribution 
is numerical resolution, as higher resolution simulations tend to produce higher eccentricities with less scatter 
\citep[see Figure 12 in][]{2017ApJ...840...53R}.
Finally, the interaction of circumbinary gas will likely affect
the eccentricity evolution \citep[e.g.][]{2011MNRAS.415.3033R},
with this effect being particularly important for somewhat lower mass SMBHs 
$(M_{\bullet} \lesssim 10^{8} M_{\sun}$).

\subsection{Relevance for Pulsar Timing Array Predictions}

It is important to account for the SMBH binary eccentricity when computing 
predictions for PTA observations. In older work this was not typically done, and instead the binaries were 
assumed to be on circular orbits \citep[e.g.][]{2008MNRAS.390..192S,kelley2017}. However, more recent treatments 
have accounted for this in the case of a fixed initial eccentricity as well as other
effects such as stellar scattering and drag from a circumbinary accretion disk
using semi-analytic models \citep[e.g.][]{2017MNRAS.471.4508K,2018ApJ...856...42S}.

These semi-analytic models appear to work quite well for the mergers considered
in this paper, as far as the processes modeled by \ketju{} are concerned.
The differences in the evolution of orbital parameters 
(see Figures \ref{fig:a_evolution} and \ref{fig:e_freq_plots}) and the GW spectrum in the PTA frequency range 
(see \autoref{fig:dEdf}) are brought down to under 5\% provided a correct choice of semi-analytic model parameters, 
and even when using incorrect parameter values or ignoring the stellar scattering completely the differences remain typically at a
level of few tens of percent.

We find that the measured hardening parameters of the 
considered semi-analytic scattering model can be smaller for our \ketju{} cored galaxy simulations  
than common literature values by factors up to 8. However, this is partially offset by the \PN{3.5}
accurate equations of motion also speeding up the merger process slightly
compared to models including only the leading quadrupole radiation reaction.
The difference to literature values is also smaller for the simulation without a stellar core 
that has less extreme SMBH masses (run X), suggesting that for most SMBH binaries the commonly
used semi-analytic models and their parameters are reasonably accurate.

Comparing the eccentricities seen in our simulations to the results of 
\cite{2017MNRAS.471.4508K}, we find that while our eccentricities are among the
highest considered there they are not quite high enough to cause the strong
suppression of the GWB seen for their model with an initial eccentricity of $e\sim 0.99$.

For PTA observations the uncertainties for the contribution from very massive binaries are probably 
dominated by discretization effects, as there are likely to be only a few very massive
systems in the final stages of a merger, with high enough GW frequencies to be
observable.  These effects affect also lower mass binaries,
and for a realistic population of SMBHs, discretization effects tend to become significant at frequencies
above $f \sim \SI{0.3}{yr^{-1}}$, depending on the eccentricities of the binaries \citep{2017MNRAS.471.4508K}.

Whether the error caused by the simplified semi-analytic models leads only to a
slightly increased statistical uncertainty, or if some kind of systematic bias
is introduced cannot be answered with our limited sample of runs. However, one should 
keep in mind that the good agreement between the
semi-analytic models and the full \ketju{} calculation was achieved when using
the properties of the stellar population derived directly from the \ketju{} runs.
Thus the strong effects of the SMBH binary on the stellar population are
accounted for in the density and velocity dispersion, while in the typical use case of semi-analytic models they are not, as
this regime falls below the gravitational softening length. This issue will particularly important 
for cosmological simulations in which the stellar population cannot be resolved at high 
spatial accuracy.

At high gravitational wave frequencies affected by discretization effects, the effects
of higher order PN-corrections can also become significant.
This is due to the fact that the GW spectrum is determined by the instantaneous
GW flux when there are only a few sources, and as was seen the difference
between the PN-accurate instantaneous flux and the commonly used quadrupole
approximation is of the order of 30\% at $f\sim \SI{1}{yr^{-1}}$ for $M \sim 10^{10} M_\sun$ binaries (see Fig \ref{fig:GW_flux}).
Ignoring these effects can thus lead to a systematic bias in the computed
background amplitude, if there are enough massive binaries for which this effect is
at its strongest.
However, we note that the corrections to the GW flux can also be handled analytically to at least 
\PN{3} order \citep[e.g.][]{2008PhRvD..77f4035A,2014LRR....17....2B}.

\subsection{Potential simulation caveats}

The derived differences between the accurate \ketju{} computations and 
the simpler semi-analytical models are relatively small, on the level of $\sim 10\%$. Thus, it is 
important to study various potential caveats of the numerical simulations, in order to assess whether
differences on this level can be reliably determined. Firstly, the stellar environment is not 
actually resolved, with each stellar particle representing about $\sim 10^5$ real stars.
Secondly, the motion of the binary is modeled only approximately using 
\PN{3.5} level corrections for the binary only, and while the accuracy of this modeling is 
state of the art for the post-Newtonian approach, the actual physics may differ from this.

The hybrid nature of \ketju{} allows us to utilize a larger number of particles
than in standard N-body codes. However, despite this fact, the actual number of stellar particles
interacting with the SMBH binary in its final stages before the merger is still
small compared to the actual number of stars that they represent.
When the SMBH binary merges in our simulations, there are typically only a few stellar particles left within \SI{10}{pc} of the binary instead of tens or hundreds of thousands.
As a result, strong interactions between the binary and stars happen far more
rarely, and when they do happen they are too strong, leading to step-like changes in the binary orbital parameters.
Fortunately, it appears that most of the environmental effects are caused by fairly weak long range 
interactions, thus the error in modeling the strong interactions should not seriously affect the long term binary evolution.

The accuracy of the binary motion also depends on the accuracy of the \PN{3.5} equations of motion.
Here we are limited by the fact that higher order equations of motion are unavailable for general binaries.
Such higher order corrections would be further complicated by their dependence on the history of the binary in addition to its instantaneous state, making implementing them impractical. 

While formally the higher order corrections should be of order $\order{\epsPN^4}$, in practice the effects may be significant even for fairly small $\epsPN$.
This was illustrated by the rather large change in the spectrum when going from
\PN{3} to \PN{3.5} equations of motion in \autoref{fig:dEdf_comparison}, even though formally the effect is of order $\epsPN^{3.5} \lesssim 10^{-3}$ there. 
The apparently fairly significant contributions from the unknown higher order 
corrections are also seen in the failure of energy conservation when the binary is very near merger (see Fig. \ref{fig:energy_conservation}). 
Since the eccentricity of the binary is fairly small in this phase, results computed for circular
binaries can be used to estimate just how much the higher order corrections affect the evolution of the binary.
According to \citet{1995PhRvL..74.3515B}, the non-linear tail-terms that are the leading correction neglected here
cause the number of orbits an equal-mass binary completes during its 
last decade in orbital frequency to change by almost 10\%,
which can be considered to be a fairly substantial effect.

\section{Conclusions}
\label{sec:conclusions}

In this paper we have studied the final stages of SMBH binary evolution in
simulations of mergers of gas-poor massive early-type galaxies using the
hybrid \ketju{} code \citep{2017ApJ...840...53R,2018ApJ...864..113R}.
In general, the evolution of the SMBH binaries was found to agree well with predictions from simple
semi-analytic models commonly used for studying gravitational wave emission from
merging SMBHs in a cosmological setting.

Differences in the emitted energy in gravitational waves 
between the \ketju{} runs and the semi-analytic models were found to be of the order $10\%$
in the frequency range of $(f\gtrsim \SI{e-1}{yr^{-1}})$, which is accessible by PTA. 
This relatively good agreement was reached provided that the properties
of the stellar population accounting for the ejection of stars via interactions with
the SMBHs and the binary eccentricity was chosen correctly.
The agreement between the semi-analytic models and the \ketju{} runs can be further improved to the level of a few percent
by using parameter values derived from the \ketju{} runs, showing that the model itself captures the relevant dynamics well.

Overall, the derived parameters from the \ketju{} runs for the semi-analytic scattering model were
found to be somewhat smaller than the literature values commonly used \citep{2006ApJ...651..392S}. 
However, this is largely explained by the fact that our runs A--D describe very massive early-type 
galaxies with large cores scoured almost empty by very massive SMBH binaries. The simulation without a 
stellar core and much smaller SMBHs (run X) is in much better agreement 
with the semi-analytic scattering model.
Determining how the effects of the SMBH binary on the stellar population affect the scattering model
parameters in general would allow reducing the error of the semi-analytic models.

The eccentricities of all of our simulated SMBH binaries were found to be very high, in excess 
of $e > 0.9$, at the onset of the gravitational wave dominated phase, when the binaries were separated by 
about $a \sim 5000 R_s$. Thus, based on our simulations the SMBHs enter the inspiral phase  
on strongly eccentric orbits, with circular orbits being viable only at the very final stages of the orbit after the 
completion of a GW driven circularization process (e.g. \citealt{2008MNRAS.390..192S,kelley2017,2017MNRAS.471.4508K}).  

In addition, the high eccentricities also result in the merger times of the binaries
being quite sensitive to small changes in their initial parameters. This is in particular the case for the 
semi-analytic models as in general the eccentricity has to be fixed at a relatively large binary separation,  
often leading to differences of some tens of percents in the binary evolution when compared to the resolved 
\ketju{} runs. 

Based on our sample of \ketju{} mergers presented in this paper and also in earlier work
\citep{2017ApJ...840...53R,2018ApJ...864..113R,2019ApJ...872L..17R} it seems that our binaries do not reach 
extremely high eccentricities in excess of $e\gtrsim 0.99$, 
which could lead to significant suppression of the
gravitational wave background amplitude detectable with PTA, as discussed in \citet{2017MNRAS.471.4508K}. 
However, a larger sample of mergers, also including cosmological simulations  would be required to 
properly characterize the SMBH eccentricity distribution at the onset of the GW driven inspiral phase.

\acknowledgments
{
The numerical simulations were performed on facilities hosted by 
the CSC -- IT Center for Science, Finland.
M.M. acknowledges the financial support by the Jenny and Antti Wihuri Foundation.
M.M., P.H.J., P.P. and A.R. acknowledge the support 
by the European Research Council via ERC Consolidator Grant KETJU (no. 818930). 
}

\software{
\ketju{} \citep{2017ApJ...840...53R,2018ApJ...864..113R},
\gadget{} \citep{Springel2005},
NumPy \citep{numpy},
SciPy \citep{scipy},
Matplotlib \citep{Hunter:2007}
}

\bibliographystyle{aasjournal}
\bibliography{refs}

\begin{thebibliography}{}
\expandafter\ifx\csname natexlab\endcsname\relax\def\natexlab#1{#1}\fi
\providecommand{\url}[1]{\href{#1}{#1}}
\providecommand{\dodoi}[1]{doi:~\href{http://doi.org/#1}{\nolinkurl{#1}}}
\providecommand{\doeprint}[1]{\href{http://ascl.net/#1}{\nolinkurl{http://ascl.net/#1}}}
\providecommand{\doarXiv}[1]{\href{https://arxiv.org/abs/#1}{\nolinkurl{https://arxiv.org/abs/#1}}}

\bibitem[{{Abbott} {et~al.}(2016)}]{Abbott2016}
{Abbott}, B.~P., {et~al.} 2016, Physical Review Letters, 116, 061102,
  \dodoi{10.1103/PhysRevLett.116.061102}

\bibitem[{{Amaro-Seoane} {et~al.}(2017){Amaro-Seoane}, {Audley}, {Babak},
  {Baker}, {Barausse}, {Bender}, {Berti}, {Binetruy}, {Born}, {Bortoluzzi},
  {Camp}, {Caprini}, {Cardoso}, {Colpi}, {Conklin}, {Cornish}, {Cutler},
  {Danzmann}, {Dolesi}, {Ferraioli}, {Ferroni}, {Fitzsimons}, {Gair}, {Gesa
  Bote}, {Giardini}, {Gibert}, {Grimani}, {Halloin}, {Heinzel}, {Hertog},
  {Hewitson}, {Holley-Bockelmann}, {Hollington}, {Hueller}, {Inchauspe},
  {Jetzer}, {Karnesis}, {Killow}, {Klein}, {Klipstein}, {Korsakova}, {Larson},
  {Livas}, {Lloro}, {Man}, {Mance}, {Martino}, {Mateos}, {McKenzie},
  {McWilliams}, {Miller}, {Mueller}, {Nardini}, {Nelemans}, {Nofrarias},
  {Petiteau}, {Pivato}, {Plagnol}, {Porter}, {Reiche}, {Robertson},
  {Robertson}, {Rossi}, {Russano}, {Schutz}, {Sesana}, {Shoemaker}, {Slutsky},
  {Sopuerta}, {Sumner}, {Tamanini}, {Thorpe}, {Troebs}, {Vallisneri},
  {Vecchio}, {Vetrugno}, {Vitale}, {Volonteri}, {Wanner}, {Ward}, {Wass},
  {Weber}, {Ziemer}, \& {Zweifel}}]{2017arXiv170200786A}
{Amaro-Seoane}, P., {Audley}, H., {Babak}, S., {et~al.} 2017, arXiv e-prints,
  arXiv:1702.00786.
\newblock \doarXiv{1702.00786}

\bibitem[{{Arun} {et~al.}(2008){Arun}, {Blanchet}, {Iyer}, \&
  {Qusailah}}]{2008PhRvD..77f4035A}
{Arun}, K.~G., {Blanchet}, L., {Iyer}, B.~R., \& {Qusailah}, M.~S.~S. 2008,
  \prd, 77, 064035, \dodoi{10.1103/PhysRevD.77.064035}

\bibitem[{{Arzoumanian} {et~al.}(2018){Arzoumanian}, {Baker}, {Brazier},
  {Burke-Spolaor}, {Chamberlin}, {Chatterjee}, {Christy}, {Cordes}, {Cornish},
  {Crawford}, {Thankful Cromartie}, {Crowter}, {DeCesar}, {Demorest}, {Dolch},
  {Ellis}, {Ferdman}, {Ferrara}, {Folkner}, {Fonseca}, {Garver-Daniels},
  {Gentile}, {Haas}, {Hazboun}, {Huerta}, {Islo}, {Jones}, {Jones}, {Kaplan},
  {Kaspi}, {Lam}, {Lazio}, {Levin}, {Lommen}, {Lorimer}, {Luo}, {Lynch},
  {Madison}, {McLaughlin}, {McWilliams}, {Mingarelli}, {Ng}, {Nice}, {Park},
  {Pennucci}, {Pol}, {Ransom}, {Ray}, {Rasskazov}, {Siemens}, {Simon},
  {Spiewak}, {Stairs}, {Stinebring}, {Stovall}, {Swiggum}, {Taylor},
  {Vallisneri}, {van Haasteren}, {Vigeland }, {Zhu}, \& {NANOGrav
  Collaboration}}]{2018ApJ...859...47A}
{Arzoumanian}, Z., {Baker}, P.~T., {Brazier}, A., {et~al.} 2018, \apj, 859, 47,
  \dodoi{10.3847/1538-4357/aabd3b}

\bibitem[{{Bansal} {et~al.}(2017){Bansal}, {Taylor}, {Peck}, {Zavala}, \&
  {Romani}}]{2017ApJ...843...14B}
{Bansal}, K., {Taylor}, G.~B., {Peck}, A.~B., {Zavala}, R.~T., \& {Romani},
  R.~W. 2017, \apj, 843, 14, \dodoi{10.3847/1538-4357/aa74e1}

\bibitem[{{Begelman} {et~al.}(1980){Begelman}, {Blandford}, \&
  {Rees}}]{begelman1980}
{Begelman}, M.~C., {Blandford}, R.~D., \& {Rees}, M.~J. 1980, \nat, 287, 307,
  \dodoi{10.1038/287307a0}

\bibitem[{{Berczik} {et~al.}(2006){Berczik}, {Merritt}, {Spurzem}, \&
  {Bischof}}]{berczik2006}
{Berczik}, P., {Merritt}, D., {Spurzem}, R., \& {Bischof}, H.-P. 2006, \apjl,
  642, L21, \dodoi{10.1086/504426}

\bibitem[{{Berentzen} {et~al.}(2009){Berentzen}, {Preto}, {Berczik}, {Merritt},
  \& {Spurzem}}]{2009ApJ...695..455B}
{Berentzen}, I., {Preto}, M., {Berczik}, P., {Merritt}, D., \& {Spurzem}, R.
  2009, \apj, 695, 455, \dodoi{10.1088/0004-637X/695/1/455}

\bibitem[{{Bernard} {et~al.}(2018){Bernard}, {Blanchet}, {Faye}, \& {Marchand
  }}]{2018PhRvD..97d4037B}
{Bernard}, L., {Blanchet}, L., {Faye}, G., \& {Marchand }, T. 2018, \prd, 97,
  044037, \dodoi{10.1103/PhysRevD.97.044037}

\bibitem[{{Binney} \& {Tremaine}(2008)}]{binney2008}
{Binney}, J., \& {Tremaine}, S. 2008, {Galactic Dynamics: Second Edition}
  (Princeton University Press)

\bibitem[{{Blanchet}(2014)}]{2014LRR....17....2B}
{Blanchet}, L. 2014, Living Reviews in Relativity, 17, 2,
  \dodoi{10.12942/lrr-2014-2}

\bibitem[{{Blanchet} {et~al.}(1995){Blanchet}, {Damour}, {Iyer}, {Will}, \&
  {Wiseman}}]{1995PhRvL..74.3515B}
{Blanchet}, L., {Damour}, T., {Iyer}, B.~R., {Will}, C.~M., \& {Wiseman}, A.~G.
  1995, Physical Review Letters, 74, 3515, \dodoi{10.1103/PhysRevLett.74.3515}

\bibitem[{{Bonetti} {et~al.}(2016){Bonetti}, {Haardt}, {Sesana}, \&
  {Barausse}}]{2016MNRAS.461.4419B}
{Bonetti}, M., {Haardt}, F., {Sesana}, A., \& {Barausse}, E. 2016, \mnras, 461,
  4419, \dodoi{10.1093/mnras/stw1590}

\bibitem[{{Bonetti} {et~al.}(2018){Bonetti}, {Sesana}, {Barausse}, \&
  {Haardt}}]{2018MNRAS.477.2599B}
{Bonetti}, M., {Sesana}, A., {Barausse}, E., \& {Haardt}, F. 2018, \mnras, 477,
  2599, \dodoi{10.1093/mnras/sty874}

\bibitem[{Bulirsch \& Stoer(1966)}]{Bulirsch1966}
Bulirsch, R., \& Stoer, J. 1966, Numerische Mathematik, 8, 1,
  \dodoi{10.1007/BF02165234}

\bibitem[{{Burke-Spolaor} {et~al.}(2019){Burke-Spolaor}, {Taylor}, {Charisi},
  {Dolch}, {Hazboun}, {Holgado}, {Kelley}, {Lazio}, {Madison}, {McMann},
  {Mingarelli}, {Rasskazov}, {Siemens}, {Simon}, \&
  {Smith}}]{2019A&ARv..27....5B}
{Burke-Spolaor}, S., {Taylor}, S.~R., {Charisi}, M., {et~al.} 2019, \aapr, 27,
  5, \dodoi{10.1007/s00159-019-0115-7}

\bibitem[{{Chapon} {et~al.}(2013){Chapon}, {Mayer}, \& {Teyssier}}]{Chapon2013}
{Chapon}, D., {Mayer}, L., \& {Teyssier}, R. 2013, \mnras, 429, 3114,
  \dodoi{10.1093/mnras/sts568}

\bibitem[{{Chen} {et~al.}(2019){Chen}, {Sesana}, \&
  {Conselice}}]{2019MNRAS.488..401C}
{Chen}, S., {Sesana}, A., \& {Conselice}, C.~J. 2019, \mnras, 488, 401,
  \dodoi{10.1093/mnras/stz1722}

\bibitem[{{Csizmadia} {et~al.}(2012){Csizmadia}, {Debreczeni}, {R{\'a}cz}, \&
  {Vas{\'u}th}}]{2012CQGra..29x5002C}
{Csizmadia}, P., {Debreczeni}, G., {R{\'a}cz}, I., \& {Vas{\'u}th}, M. 2012,
  Classical and Quantum Gravity, 29, 245002,
  \dodoi{10.1088/0264-9381/29/24/245002}

\bibitem[{{Deane} {et~al.}(2014){Deane}, {Paragi}, {Jarvis}, {Coriat},
  {Bernardi}, {Fender}, {Frey}, {Heywood}, {Kl{\"o}ckner}, {Grainge}, \&
  {Rumsey}}]{2014Natur.511...57D}
{Deane}, R.~P., {Paragi}, Z., {Jarvis}, M.~J., {et~al.} 2014, \nat, 511, 57,
  \dodoi{10.1038/nature13454}

\bibitem[{{Dehnen}(1993)}]{Dehnen1993}
{Dehnen}, W. 1993, \mnras, 265, 250, \dodoi{10.1093/mnras/265.1.250}

\bibitem[{{Dey} {et~al.}(2018){Dey}, {Valtonen}, {Gopakumar}, {Zola}, {Hudec},
  {Pihajoki}, {Ciprini}, {Matsumoto}, {Sadakane}, {Kidger}, {Nilsson},
  {Mikkola}, {Sillanp{\"a}{\"a}}, {Takalo}, {Lehto}, {Berdyugin}, {Piirola},
  {Jermak}, {Baliyan}, {Pursimo}, {Caton}, {Alicavus}, {Baransky}, {Blay},
  {Boumis}, {Boyd}, {Campas Torrent}, {Campos}, {Carrillo G{\'o}mez},
  {Chandra}, {Chavushyan}, {Dalessio}, {Debski}, {Drozdz}, {Er}, {Erdem},
  {Escartin P{\'e}rez}, {Fallah Ramazani}, {Filippenko}, {Gafton}, {Ganesh},
  {Garcia}, {Gazeas}, {Godunova}, {G{\'o}mez Pinilla}, {Gopinathan}, {Haislip},
  {Harmanen}, {Hurst}, {Jan{\'{\i}}k}, {Jelinek}, {Joshi}, {Kagitani},
  {Karjalainen}, {Kaur}, {Keel}, {Kouprianov}, {Kundera}, {Kurowski},
  {Kvammen}, {LaCluyze}, {Lee}, {Liakos}, {Lindfors}, {Lozano de Haro},
  {Mugrauer}, {Naves Nogues}, {Neely}, {Nelson}, {Ogloza}, {Okano},
  {Pajdosz-{\'S}mierciak}, {Pandey}, {Perri}, {Poyner}, {Provencal}, {Raj},
  {Reichart}, {Reinthal}, {Reynolds}, {Saario}, {Sadegi}, {Sakanoi}, {Salto
  Gonz{\'a}lez}, {Sameer}, {Schweyer}, {Simon}, {Siwak}, {Sold{\'a}n Alfaro},
  {Sonbas}, {Steele}, {Stocke}, {Strobl}, {Tomov}, {Tremosa Espasa}, {Valdes},
  {Valero P{\'e}rez}, {Verrecchia}, {Vasylenko}, {Webb}, {Yoneda}, {Zejmo},
  {Zheng}, \& {Zielinski}}]{2018ApJ...866...11D}
{Dey}, L., {Valtonen}, M.~J., {Gopakumar}, A., {et~al.} 2018, \apj, 866, 11,
  \dodoi{10.3847/1538-4357/aadd95}

\bibitem[{{Eisenreich} {et~al.}(2017){Eisenreich}, {Naab}, {Choi}, {Ostriker},
  \& {Emsellem}}]{Eisenreich2017}
{Eisenreich}, M., {Naab}, T., {Choi}, E., {Ostriker}, J.~P., \& {Emsellem}, E.
  2017, \mnras, 468, 751, \dodoi{10.1093/mnras/stx473}

\bibitem[{{Enoki} \& {Nagashima}(2007)}]{2007PThPh.117..241E}
{Enoki}, M., \& {Nagashima}, M. 2007, Progress of Theoretical Physics, 117,
  241, \dodoi{10.1143/PTP.117.241}

\bibitem[{{Furlong} {et~al.}(2017){Furlong}, {Bower}, {Crain}, {Schaye},
  {Theuns}, {Trayford}, {Qu}, {Schaller}, {Berthet}, \&
  {Helly}}]{2017MNRAS.465..722F}
{Furlong}, M., {Bower}, R.~G., {Crain}, R.~A., {et~al.} 2017, \mnras, 465, 722,
  \dodoi{10.1093/mnras/stw2740}

\bibitem[{{Gragg}(1965)}]{Gragg1965}
{Gragg}, W.~B. 1965, SIAM Journal on Numerical Analysis, 2, 384,
  \dodoi{10.1137/0702030}

\bibitem[{{Gualandris} {et~al.}(2017){Gualandris}, {Read}, {Dehnen}, \&
  {Bortolas}}]{2017MNRAS.464.2301G}
{Gualandris}, A., {Read}, J.~I., {Dehnen}, W., \& {Bortolas}, E. 2017, \mnras,
  464, 2301, \dodoi{10.1093/mnras/stw2528}

\bibitem[{{Hellstr{\"o}m} \& {Mikkola}(2010)}]{Hellstrom2010}
{Hellstr{\"o}m}, C., \& {Mikkola}, S. 2010, Celestial Mechanics and Dynamical
  Astronomy, 106, 143, \dodoi{10.1007/s10569-009-9248-8}

\bibitem[{{Hernquist}(1990)}]{hernquist1990}
{Hernquist}, L. 1990, \apj, 356, 359, \dodoi{10.1086/168845}

\bibitem[{{Hills} \& {Fullerton}(1980)}]{1980AJ.....85.1281H}
{Hills}, J.~G., \& {Fullerton}, L.~W. 1980, \aj, 85, 1281,
  \dodoi{10.1086/112798}

\bibitem[{{Hilz} {et~al.}(2012){Hilz}, {Naab}, {Ostriker}, {Thomas}, {Burkert},
  \& {Jesseit}}]{Hilz2012}
{Hilz}, M., {Naab}, T., {Ostriker}, J.~P., {et~al.} 2012, \mnras, 425, 3119,
  \dodoi{10.1111/j.1365-2966.2012.21541.x}

\bibitem[{{Huerta} {et~al.}(2015){Huerta}, {McWilliams}, {Gair}, \&
  {Taylor}}]{2015PhRvD..92f3010H}
{Huerta}, E.~A., {McWilliams}, S.~T., {Gair}, J.~R., \& {Taylor}, S.~R. 2015,
  \prd, 92, 063010, \dodoi{10.1103/PhysRevD.92.063010}

\bibitem[{Hunter(2007)}]{Hunter:2007}
Hunter, J.~D. 2007, Computing in Science \& Engineering, 9, 90,
  \dodoi{10.1109/MCSE.2007.55}

\bibitem[{{Inayoshi} {et~al.}(2018){Inayoshi}, {Ichikawa}, \&
  {Haiman}}]{2018ApJ...863L..36I}
{Inayoshi}, K., {Ichikawa}, K., \& {Haiman}, Z. 2018, \apjl, 863, L36,
  \dodoi{10.3847/2041-8213/aad8ad}

\bibitem[{{Johansson} {et~al.}(2012){Johansson}, {Naab}, \&
  {Ostriker}}]{2012Johansson}
{Johansson}, P.~H., {Naab}, T., \& {Ostriker}, J.~P. 2012, \apj, 754, 115,
  \dodoi{10.1088/0004-637X/754/2/115}

\bibitem[{Jones {et~al.}(2001--)Jones, Oliphant, Peterson, {et~al.}}]{scipy}
Jones, E., Oliphant, T., Peterson, P., {et~al.} 2001--, {SciPy}: Open source
  scientific tools for {Python}.
\newblock \url{http://www.scipy.org/}

\bibitem[{{Karl} {et~al.}(2015){Karl}, {Aarseth}, {Naab}, {Haehnelt}, \&
  {Spurzem}}]{2015MNRAS.452.2337K}
{Karl}, S.~J., {Aarseth}, S.~J., {Naab}, T., {Haehnelt}, M.~G., \& {Spurzem},
  R. 2015, \mnras, 452, 2337, \dodoi{10.1093/mnras/stv1453}

\bibitem[{{Kelley} {et~al.}(2017{\natexlab{a}}){Kelley}, {Blecha}, \&
  {Hernquist}}]{kelley2017}
{Kelley}, L.~Z., {Blecha}, L., \& {Hernquist}, L. 2017{\natexlab{a}}, \mnras,
  464, 3131, \dodoi{10.1093/mnras/stw2452}

\bibitem[{{Kelley} {et~al.}(2017{\natexlab{b}}){Kelley}, {Blecha}, {Hernquist},
  {Sesana}, \& {Taylor}}]{2017MNRAS.471.4508K}
{Kelley}, L.~Z., {Blecha}, L., {Hernquist}, L., {Sesana}, A., \& {Taylor},
  S.~R. 2017{\natexlab{b}}, \mnras, 471, 4508, \dodoi{10.1093/mnras/stx1638}

\bibitem[{{Khan} {et~al.}(2018{\natexlab{a}}){Khan}, {Berczik}, \&
  {Just}}]{2018A&A...615A..71K}
{Khan}, F.~M., {Berczik}, P., \& {Just}, A. 2018{\natexlab{a}}, \aap, 615, A71,
  \dodoi{10.1051/0004-6361/201730489}

\bibitem[{{Khan} {et~al.}(2018{\natexlab{b}}){Khan}, {Capelo}, {Mayer}, \&
  {Berczik}}]{2018ApJ...868...97K}
{Khan}, F.~M., {Capelo}, P.~R., {Mayer}, L., \& {Berczik}, P.
  2018{\natexlab{b}}, \apj, 868, 97, \dodoi{10.3847/1538-4357/aae77b}

\bibitem[{{Khan} {et~al.}(2016){Khan}, {Fiacconi}, {Mayer}, {Berczik}, \&
  {Just}}]{2016ApJ...828...73K}
{Khan}, F.~M., {Fiacconi}, D., {Mayer}, L., {Berczik}, P., \& {Just}, A. 2016,
  \apj, 828, 73, \dodoi{10.3847/0004-637X/828/2/73}

\bibitem[{{Khan} {et~al.}(2013){Khan}, {Holley-Bockelmann}, {Berczik}, \&
  {Just}}]{2013ApJ...773..100K}
{Khan}, F.~M., {Holley-Bockelmann}, K., {Berczik}, P., \& {Just}, A. 2013,
  \apj, 773, 100, \dodoi{10.1088/0004-637X/773/2/100}

\bibitem[{{Khan} {et~al.}(2011){Khan}, {Just}, \&
  {Merritt}}]{2011ApJ...732...89K}
{Khan}, F.~M., {Just}, A., \& {Merritt}, D. 2011, \apj, 732, 89,
  \dodoi{10.1088/0004-637X/732/2/89}

\bibitem[{{Khochfar} \& {Burkert}(2006)}]{Khochfar2006}
{Khochfar}, S., \& {Burkert}, A. 2006, \aap, 445, 403,
  \dodoi{10.1051/0004-6361:20053241}

\bibitem[{{Klein} {et~al.}(2018){Klein}, {Boetzel}, {Gopakumar}, {Jetzer}, \&
  {de Vittori}}]{2018PhRvD..98j4043K}
{Klein}, A., {Boetzel}, Y., {Gopakumar}, A., {Jetzer}, P., \& {de Vittori}, L.
  2018, \prd, 98, 104043, \dodoi{10.1103/PhysRevD.98.104043}

\bibitem[{{K{\"o}nigsd{\"o}rffer} \& {Gopakumar}(2006)}]{2006PhRvD..73l4012K}
{K{\"o}nigsd{\"o}rffer}, C., \& {Gopakumar}, A. 2006, \prd, 73, 124012,
  \dodoi{10.1103/PhysRevD.73.124012}

\bibitem[{{Kormendy} \& {Ho}(2013)}]{kormendy2013}
{Kormendy}, J., \& {Ho}, L.~C. 2013, \araa, 51, 511,
  \dodoi{10.1146/annurev-astro-082708-101811}

\bibitem[{{Mayer} {et~al.}(2007){Mayer}, {Kazantzidis}, {Madau}, {Colpi},
  {Quinn}, \& {Wadsley}}]{Mayer2007}
{Mayer}, L., {Kazantzidis}, S., {Madau}, P., {et~al.} 2007, Science, 316, 1874,
  \dodoi{10.1126/science.1141858}

\bibitem[{{McWilliams} {et~al.}(2014){McWilliams}, {Ostriker}, \&
  {Pretorius}}]{2014ApJ...789..156M}
{McWilliams}, S.~T., {Ostriker}, J.~P., \& {Pretorius}, F. 2014, \apj, 789,
  156, \dodoi{10.1088/0004-637X/789/2/156}

\bibitem[{{Memmesheimer} {et~al.}(2004){Memmesheimer}, {Gopakumar}, \&
  {Sch{\"a}fer}}]{2004PhRvD..70j4011M}
{Memmesheimer}, R.-M., {Gopakumar}, A., \& {Sch{\"a}fer}, G. 2004, \prd, 70,
  104011, \dodoi{10.1103/PhysRevD.70.104011}

\bibitem[{{Merritt}(1985)}]{Merritt1985}
{Merritt}, D. 1985, \aj, 90, 1027, \dodoi{10.1086/113810}

\bibitem[{Merritt(2013)}]{merritt2013}
Merritt, D. 2013, Dynamics and evolution of galactic nuclei (Princeton
  University Press)

\bibitem[{{Middleton} {et~al.}(2018){Middleton}, {Chen}, {Del Pozzo}, {Sesana},
  \& {Vecchio}}]{2018NatCo...9..573M}
{Middleton}, H., {Chen}, S., {Del Pozzo}, W., {Sesana}, A., \& {Vecchio}, A.
  2018, Nature Communications, 9, 573, \dodoi{10.1038/s41467-018-02916-7}

\bibitem[{{Mikkola} \& {Aarseth}(1993)}]{Mikkola1993}
{Mikkola}, S., \& {Aarseth}, S.~J. 1993, Celestial Mechanics and Dynamical
  Astronomy, 57, 439, \dodoi{10.1007/BF00695714}

\bibitem[{{Mikkola} \& {Merritt}(2006)}]{Mikkola2006}
{Mikkola}, S., \& {Merritt}, D. 2006, \mnras, 372, 219,
  \dodoi{10.1111/j.1365-2966.2006.10854.x}

\bibitem[{{Mikkola} \& {Merritt}(2008)}]{Mikkola2008}
---. 2008, \aj, 135, 2398, \dodoi{10.1088/0004-6256/135/6/2398}

\bibitem[{{Mikkola} \& {Tanikawa}(1999)}]{1999MNRAS.310..745M}
{Mikkola}, S., \& {Tanikawa}, K. 1999, \mnras, 310, 745,
  \dodoi{10.1046/j.1365-8711.1999.02982.x}

\bibitem[{{Milosavljevi{\'c}} \& {Merritt}(2001)}]{Milosavljevic2001}
{Milosavljevi{\'c}}, M., \& {Merritt}, D. 2001, \apj, 563, 34,
  \dodoi{10.1086/323830}

\bibitem[{{Mora} \& {Will}(2004)}]{2004PhRvD..69j4021M}
{Mora}, T., \& {Will}, C.~M. 2004, \prd, 69, 104021,
  \dodoi{10.1103/PhysRevD.69.104021}

\bibitem[{{Moster} {et~al.}(2018){Moster}, {Naab}, \&
  {White}}]{2018MNRAS.477.1822M}
{Moster}, B.~P., {Naab}, T., \& {White}, S.~D.~M. 2018, \mnras, 477, 1822,
  \dodoi{10.1093/mnras/sty655}

\bibitem[{{Naab} {et~al.}(2009){Naab}, {Johansson}, \& {Ostriker}}]{2009Naab}
{Naab}, T., {Johansson}, P.~H., \& {Ostriker}, J.~P. 2009, \apjl, 699, L178,
  \dodoi{10.1088/0004-637X/699/2/L178}

\bibitem[{{Naab} \& {Ostriker}(2017)}]{2017ARA&A..55...59N}
{Naab}, T., \& {Ostriker}, J.~P. 2017, \araa, 55, 59,
  \dodoi{10.1146/annurev-astro-081913-040019}

\bibitem[{{Oser} {et~al.}(2010){Oser}, {Ostriker}, {Naab}, {Johansson}, \&
  {Burkert}}]{oser2010}
{Oser}, L., {Ostriker}, J.~P., {Naab}, T., {Johansson}, P.~H., \& {Burkert}, A.
  2010, \apj, 725, 2312, \dodoi{10.1088/0004-637X/725/2/2312}

\bibitem[{{Peters}(1964)}]{1964PhRv..136.1224P}
{Peters}, P.~C. 1964, Physical Review, 136, 1224,
  \dodoi{10.1103/PhysRev.136.B1224}

\bibitem[{{Peters} \& {Mathews}(1963)}]{1963PhRv..131..435P}
{Peters}, P.~C., \& {Mathews}, J. 1963, Physical Review, 131, 435,
  \dodoi{10.1103/PhysRev.131.435}

\bibitem[{{Phinney}(2001)}]{2001astro.ph..8028P}
{Phinney}, E.~S. 2001, arXiv e-prints, astro.
\newblock \doarXiv{astro-ph/0108028}

\bibitem[{{Pihajoki}(2015)}]{Pihajoki2015}
{Pihajoki}, P. 2015, Celestial Mechanics and Dynamical Astronomy, 121, 211,
  \dodoi{10.1007/s10569-014-9597-9}

\bibitem[{Poisson \& Will(2014)}]{poisson2014gravity}
Poisson, E., \& Will, C.~M. 2014, Gravity: Newtonian, Post-Newtonian,
  Relativistic (Cambridge University Press), \dodoi{10.1017/CBO9781139507486}

\bibitem[{{Preto} {et~al.}(2011){Preto}, {Berentzen}, {Berczik}, \&
  {Spurzem}}]{2011ApJ...732L..26P}
{Preto}, M., {Berentzen}, I., {Berczik}, P., \& {Spurzem}, R. 2011, \apjl, 732,
  L26, \dodoi{10.1088/2041-8205/732/2/L26}

\bibitem[{{Preto} \& {Tremaine}(1999)}]{1999AJ....118.2532P}
{Preto}, M., \& {Tremaine}, S. 1999, \aj, 118, 2532, \dodoi{10.1086/301102}

\bibitem[{{Quinlan}(1996)}]{1996NewA....1...35Q}
{Quinlan}, G.~D. 1996, \na, 1, 35, \dodoi{10.1016/S1384-1076(96)00003-6}

\bibitem[{{Rantala} {et~al.}(2018){Rantala}, {Johansson}, {Naab}, {Thomas}, \&
  {Frigo}}]{2018ApJ...864..113R}
{Rantala}, A., {Johansson}, P.~H., {Naab}, T., {Thomas}, J., \& {Frigo}, M.
  2018, \apj, 864, 113, \dodoi{10.3847/1538-4357/aada47}

\bibitem[{{Rantala} {et~al.}(2019){Rantala}, {Johansson}, {Naab}, {Thomas}, \&
  {Frigo}}]{2019ApJ...872L..17R}
---. 2019, \apjl, 872, L17, \dodoi{10.3847/2041-8213/ab04b1}

\bibitem[{{Rantala} {et~al.}(2017){Rantala}, {Pihajoki}, {Johansson}, {Naab},
  {Lah{\'e}n}, \& {Sawala}}]{2017ApJ...840...53R}
{Rantala}, A., {Pihajoki}, P., {Johansson}, P.~H., {et~al.} 2017, \apj, 840,
  53, \dodoi{10.3847/1538-4357/aa6d65}

\bibitem[{{Rodriguez} {et~al.}(2006){Rodriguez}, {Taylor}, {Zavala}, {Peck},
  {Pollack}, \& {Romani}}]{2006ApJ...646...49R}
{Rodriguez}, C., {Taylor}, G.~B., {Zavala}, R.~T., {et~al.} 2006, \apj, 646,
  49, \dodoi{10.1086/504825}

\bibitem[{{Roedig} {et~al.}(2011){Roedig}, {Dotti}, {Sesana}, {Cuadra}, \&
  {Colpi}}]{2011MNRAS.415.3033R}
{Roedig}, C., {Dotti}, M., {Sesana}, A., {Cuadra}, J., \& {Colpi}, M. 2011,
  \mnras, 415, 3033, \dodoi{10.1111/j.1365-2966.2011.18927.x}

\bibitem[{{Ryu} {et~al.}(2018){Ryu}, {Perna}, {Haiman}, {Ostriker}, \&
  {Stone}}]{2018MNRAS.473.3410R}
{Ryu}, T., {Perna}, R., {Haiman}, Z., {Ostriker}, J.~P., \& {Stone}, N.~C.
  2018, \mnras, 473, 3410, \dodoi{10.1093/mnras/stx2524}

\bibitem[{{Salcido} {et~al.}(2016){Salcido}, {Bower}, {Theuns}, {McAlpine},
  {Schaller}, {Crain}, {Schaye}, \& {Regan}}]{2016MNRAS.463..870S}
{Salcido}, J., {Bower}, R.~G., {Theuns}, T., {et~al.} 2016, \mnras, 463, 870,
  \dodoi{10.1093/mnras/stw2048}

\bibitem[{{Sesana}(2010)}]{2010ApJ...719..851S}
{Sesana}, A. 2010, \apj, 719, 851, \dodoi{10.1088/0004-637X/719/1/851}

\bibitem[{{Sesana}(2013)}]{2013CQGra..30v4014S}
---. 2013, Classical and Quantum Gravity, 30, 224014,
  \dodoi{10.1088/0264-9381/30/22/224014}

\bibitem[{{Sesana} {et~al.}(2006){Sesana}, {Haardt}, \&
  {Madau}}]{2006ApJ...651..392S}
{Sesana}, A., {Haardt}, F., \& {Madau}, P. 2006, \apj, 651, 392,
  \dodoi{10.1086/507596}

\bibitem[{{Sesana} {et~al.}(2018){Sesana}, {Haiman}, {Kocsis}, \&
  {Kelley}}]{2018ApJ...856...42S}
{Sesana}, A., {Haiman}, Z., {Kocsis}, B., \& {Kelley}, L.~Z. 2018, \apj, 856,
  42, \dodoi{10.3847/1538-4357/aaad0f}

\bibitem[{{Sesana} \& {Khan}(2015)}]{2015MNRAS.454L..66S}
{Sesana}, A., \& {Khan}, F.~M. 2015, \mnras, 454, L66,
  \dodoi{10.1093/mnrasl/slv131}

\bibitem[{{Sesana} {et~al.}(2008){Sesana}, {Vecchio}, \&
  {Colacino}}]{2008MNRAS.390..192S}
{Sesana}, A., {Vecchio}, A., \& {Colacino}, C.~N. 2008, \mnras, 390, 192,
  \dodoi{10.1111/j.1365-2966.2008.13682.x}

\bibitem[{{Springel} {et~al.}(2005){Springel}, {Di Matteo}, \&
  {Hernquist}}]{Springel2005}
{Springel}, V., {Di Matteo}, T., \& {Hernquist}, L. 2005, \mnras, 361, 776,
  \dodoi{10.1111/j.1365-2966.2005.09238.x}

\bibitem[{{Tessmer} \& {Sch{\"a}fer}(2010)}]{2010PhRvD..82l4064T}
{Tessmer}, M., \& {Sch{\"a}fer}, G. 2010, \prd, 82, 124064,
  \dodoi{10.1103/PhysRevD.82.124064}

\bibitem[{{Thomas} {et~al.}(2016){Thomas}, {Ma}, {McConnell}, {Greene},
  {Blakeslee}, \& {Janish}}]{thomas2016}
{Thomas}, J., {Ma}, C.-P., {McConnell}, N.~J., {et~al.} 2016, \nat, 532, 340,
  \dodoi{10.1038/nature17197}

\bibitem[{{Tiburzi}(2018)}]{2018PASA...35...13T}
{Tiburzi}, C. 2018, \pasa, 35, e013, \dodoi{10.1017/pasa.2018.7}

\bibitem[{{Tremmel} {et~al.}(2018){Tremmel}, {Governato}, {Volonteri}, {Quinn},
  \& {Pontzen}}]{2018MNRAS.475.4967T}
{Tremmel}, M., {Governato}, F., {Volonteri}, M., {Quinn}, T.~R., \& {Pontzen},
  A. 2018, \mnras, 475, 4967, \dodoi{10.1093/mnras/sty139}

\bibitem[{{Valtonen} {et~al.}(2008){Valtonen}, {Lehto}, {Nilsson}, {Heidt},
  {Takalo}, {Sillanp{\"a}{\"a}}, {Villforth}, {Kidger}, {Poyner}, {Pursimo},
  {Zola}, {Wu}, {Zhou}, {Sadakane}, {Drozdz}, {Koziel}, {Marchev}, {Ogloza},
  {Porowski}, {Siwak}, {Stachowski}, {Winiarski}, {Hentunen}, {Nissinen},
  {Liakos}, \& {Dogru}}]{2008Natur.452..851V}
{Valtonen}, M.~J., {Lehto}, H.~J., {Nilsson}, K., {et~al.} 2008, \nat, 452,
  851, \dodoi{10.1038/nature06896}

\bibitem[{{van der Walt} {et~al.}(2011){van der Walt}, {Colbert}, \&
  {Varoquaux}}]{numpy}
{van der Walt}, S., {Colbert}, S.~C., \& {Varoquaux}, G. 2011, Computing in
  Science Engineering, 13, 22, \dodoi{10.1109/MCSE.2011.37}

\bibitem[{{Vasiliev} {et~al.}(2015){Vasiliev}, {Antonini}, \&
  {Merritt}}]{Vasiliev2015}
{Vasiliev}, E., {Antonini}, F., \& {Merritt}, D. 2015, \apj, 810, 49,
  \dodoi{10.1088/0004-637X/810/1/49}

\bibitem[{{Volonteri} {et~al.}(2003){Volonteri}, {Haardt}, \&
  {Madau}}]{volonteri2003}
{Volonteri}, M., {Haardt}, F., \& {Madau}, P. 2003, \apj, 582, 559,
  \dodoi{10.1086/344675}

\bibitem[{{Wellons} {et~al.}(2015)}]{2015Wellons}
{Wellons}, S., {et~al.} 2015, \mnras, 449, 361, \dodoi{10.1093/mnras/stv303}

\bibitem[{{Will}(2006)}]{Will2006}
{Will}, C.~M. 2006, Living Reviews in Relativity, 9, 3,
  \dodoi{10.12942/lrr-2006-3}

\end{thebibliography}

\end{document}